\documentclass[fleqn,usenatbib,onecolumn]{mnras}
\usepackage{newtxtext,newtxmath}
\usepackage[T1]{fontenc}
\usepackage{ae,aecompl}
\usepackage{graphicx}	
\usepackage{amsmath}	
\usepackage{amssymb}	
\usepackage{extarrows}

\usepackage{amsthm}
\usepackage{caption}
\usepackage{color}
\usepackage{longtable}
\usepackage{supertabular,booktabs}
\usepackage[utf8]{inputenc}
\usepackage{url}
\usepackage{multirow}
\usepackage{scrextend}
\usepackage[flushleft]{threeparttable}
\usepackage{braket}
\usepackage{mathtools, cuted}
\usepackage{enumitem}
\usepackage{floatrow}
\usepackage{float}
\DeclareMathOperator{\arccot}{arccot}


\title{
Quantum mechanics at high redshift -- Modelling Damped Lyman--$\alpha$ absorption systems}

\author[C. C. Lee et al]{
C. C. Lee$^{1}$\thanks{E-mail: lee.chungchi16@gmail.com},
J. K. Webb$^2$,
R.F. Carswell$^3$.\\ \\
$^{1}$DAMTP, Centre for Mathematical Sciences, University of Cambridge, Cambridge CB3 0WA, UK\\
$^{2}$School of Physics, University of New South Wales Sydney, NSW 2052, Australia\\
$^{3}$Institute of Astronomy, Madingley Road, Cambridge CB3 0HA, UK
}

\date{Accepted XXX. Received YYY; in original form ZZZ}
\pubyear{2019}

\begin{document}
\label{firstpage}
\maketitle

\begin{abstract}
For around 100 years, hydrogen spectral modelling has been based on Voigt profile fitting. The semi-classical Voigt profile is based on a 2-level atom approximation. Whilst the Voigt profile is excellent for many circumstances, the accuracy is insufficient for very high column density damped Lyman-$\alpha$ absorption systems. We have adapted the quantum-mechanical Kramers-Heisenberg model to include thermal broadening, producing a new profile, the KHT profile. Interactions involving multiple discrete atomic levels and continuum terms, not accounted for in the Voigt model, generate asymmetries in the Lyman line wings. If not modelled, this can lead to significant systematics in parameter estimation when modelling real data. There are important ramifications in particular for measurements of the primordial deuterium abundance. However, the KHT model is complicated. We therefore present a simplified formulation based on Taylor series expansions and look-up tables, quantifying the impact of the approximations made. The KHT profile has been implemented within the widely-used VPFIT code.

\end{abstract}

\begin{keywords}
cosmology: theory -- quasars: absorption lines -- atomic processes -- methods: data analysis -- radiation mechanisms:general -- cosmology: cosmological parameters
\end{keywords}

\section{Introduction}

Damped Lyman-$\alpha$ absorption systems (DLAs), seen towards distant quasars, offer a unique opportunity to probe the early universe. To date, DLA modelling has involved fitting Voigt profiles to sufficiently high quality spectroscopic data. For simplicity, the Voigt model treats each atomic transition in isolation as a classical 2-level damped harmonic oscillator, ignoring interactions associated with other discrete energy levels, and excluding transitions from continuum states to other bound levels (discussed later in this paper). The natural line broadening of the Voigt model, described by its Lorentzian component, is based on the Uncertainty Principle. The Voigt model, which can therefore be thought of as semi-classical, provides an excellent approximation under many circumstances. However, the pioneering work of H.-W. Lee and colleagues shows that the Voigt profile begins to deviate significantly from the fully quantum mechanical Kramers-Heisenberg model at high atomic column densities. A detailed discussion of the increasing asymmetry of the Kramers-Heisenberg profile at high column densities is given in \cite{Lee2013}.

The Kramers-Heisenberg model is considerably more complicated than the relatively simple Voigt function. Moreover, the Kramers-Heisenberg model includes only line-damping effects and ignores the effects of thermal line broadening, unlike Voigt. Including thermal broadening is essential for application to real data in certain circumstances (described later) although, unfortunately, the mathematics becomes even more cumbersome when doing so, as we shall see in this paper.

The primary three goals of this work are as follows: (i) Analogous to the necessity of adapting the Lorentzian profile by convolution with a Maxwellian function to include thermal broadening (the Voigt profile), we adapt the Kramers-Heisenberg model to produce a new profile that explicitly incorporates thermal broadening. We call this the KHT profile; (ii) In order to find practical ways of computing the KHT profile such that applying it to real data becomes feasible, we derive Taylor series approximations to the KHT integral equations; (iii) We implement the methods, using a set of look-up tables, commenting on computing speeds and other practicalities, and incorporate the KHT profile into the widely used code, VPFIT\footnote{\url{https://www.ast.cam.ac.uk/~rfc/}}.

The rest of this paper is organised as follows: in Section \ref{sec:cross} we first introduce the Kramers-Heisenberg profile, illustrating how it is based on an infinite sum over all bound levels plus an integral over all continuum states and then describe the calculation of the two (discrete and continuous) terms. In Section \ref{sec:broadening} we incorporate thermal broadening, deriving the KHT profile, a convolution of the Kramers-Heisenberg and Maxwellian profiles. In Section \ref{sec:eval_kh} we show how using look-up tables makes calculating the KHT profile practical. Finally, in Section \ref{sec:discussion} we summarise the main outcomes of this work and discuss some implications. Four appendices provide additional derivation details, illustrate how the Kramers-Heisenberg and KHT profiles reduce to Lorentzian and Voigt profiles respectively, and tabulate new damping coefficient calculations that were required in the course of this work.

\section{The Kramers-Heisenberg absorption cross-section}
\label{sec:cross}

The Kramers-Heisenberg cross-section was derived shortly before a final quantum theory had been completed \citep{Kramers1924a,Breit1924,Kramers1924b,vanVleck1924,Kramers1925}. Shortly afterwards, \citep{Dirac1927} gave the first fully quantum mechanical derivation of the Kramers-Heisenberg theory. \cite{Breit1932} gives a comprehensive review. For comparative purposes, a convenient approximation to the cross-section is given in \cite{Peebles1993}. The Kramers-Heisenberg theory has also been recently discussed in \cite{Mortlock2016}.

\begin{figure*}
\centering
\floatbox[{\capbeside\thisfloatsetup{capbesideposition={right,center},capbesidewidth=-4 cm}}]{figure}[\FBwidth]
{\includegraphics[width=1.0\linewidth]{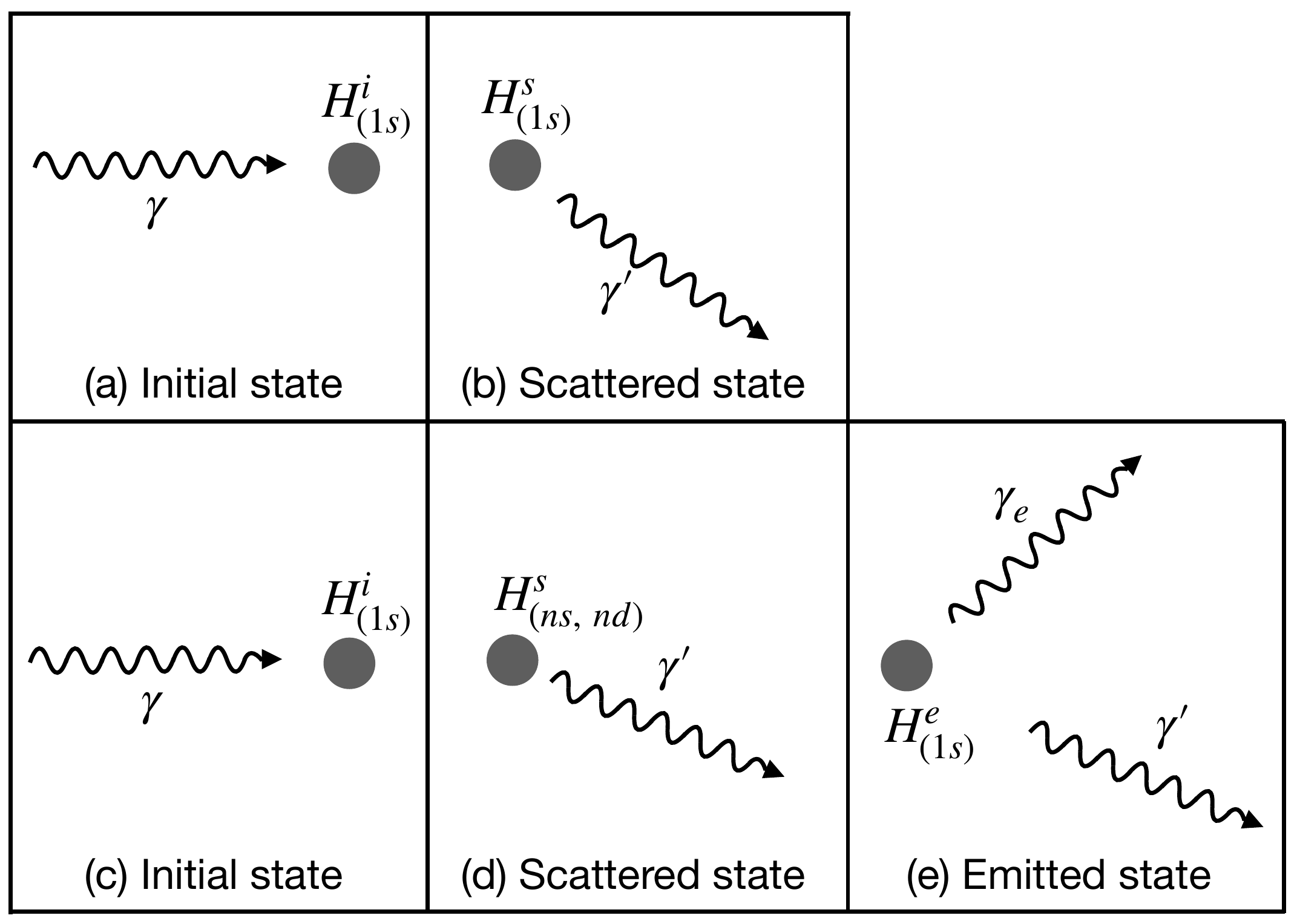}}
{\caption{. Schematic showing the physical processes associated with the total absorption cross-section for a single-nucleus atom (i.e. H or D, but designated $H$ here). Panel (a): An incoming photon, $\gamma$, collides with an atom in an initial ground-state, 1s. Panel (b): The scattered photon, $\gamma^\prime$, leaves the atom still in its ground-state, the incoming and scattered photon energies being identical. Panels (a) and (b) together illustrate Rayleigh scattering, described by Equation \eqref{eq:ray1}. Panel (c) again represents the initial 1s state. Panel (d): The scattered photon $\gamma^\prime$ leaves the atom in an excited state (ns with $n>1$, or nd). Panel (e): The atom drops back to lower states (ultimately reaching the ground state), emitting photons, $\gamma_e$, in the process. Panels (c) and (d) illustrate Raman scattering, described by Equation \eqref{eq:ram1}. Since the processes illustrated in panel (e) is emission, this is not directly relevant to the Kramers-Heisenberg calculation.}
\label{fig:scattering}}
\end{figure*}

\begin{figure*}
\centering
\floatbox[{\capbeside\thisfloatsetup{capbesideposition={right,center},capbesidewidth=-4 cm}}]{figure}[\FBwidth]
{\includegraphics[width=1.0\linewidth]{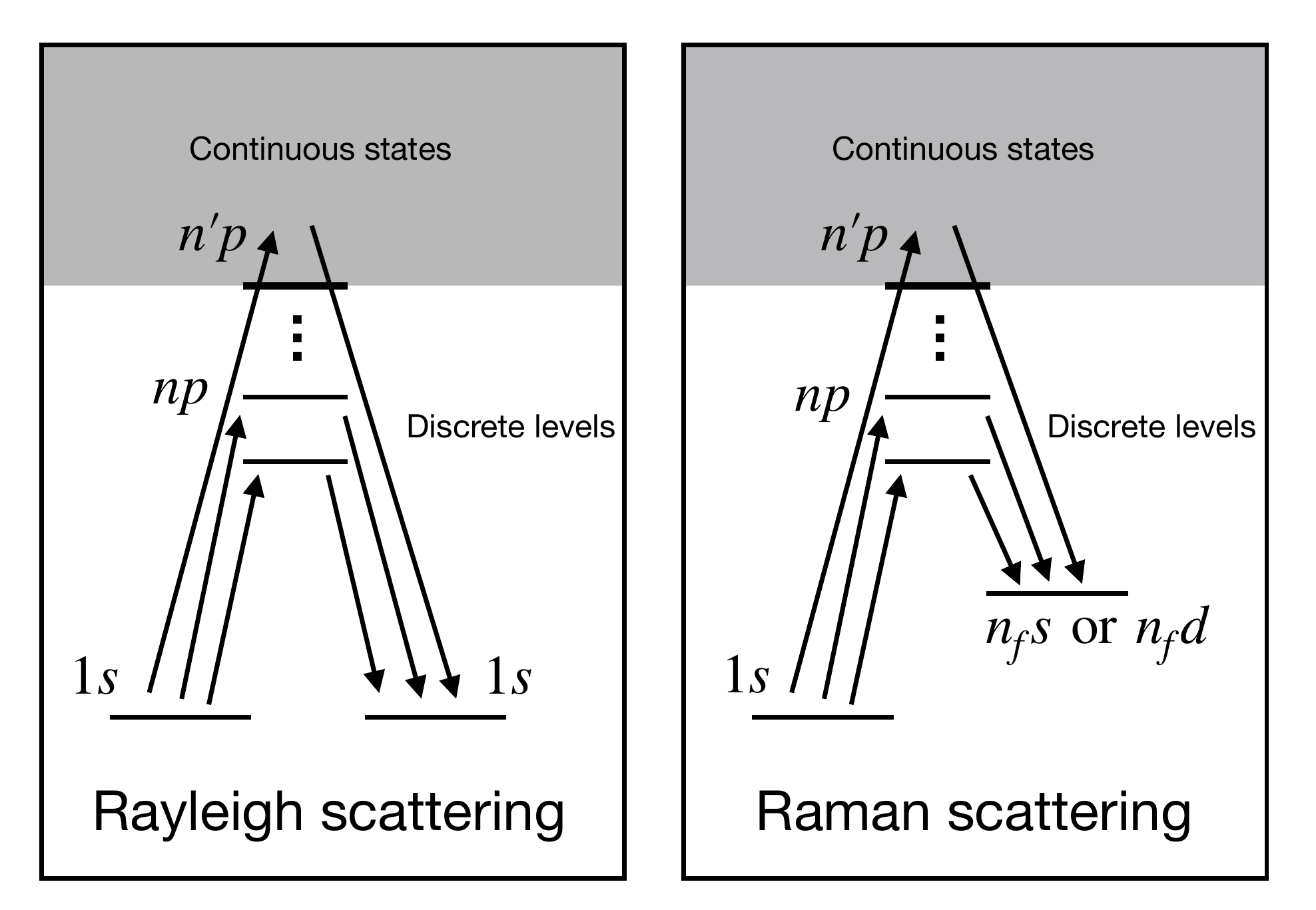}}
{\caption{. Energy level representation of the Rayleigh and Raman discrete and continuous transitions contributing to the Kramers-Heisenberg absorption cross-section. The left-hand panel corresponds to panels (a) and (b) in Figure \ref{fig:scattering}. The right-hand panel corresponds to panels (c), (d), and (e) in Figure \ref{fig:scattering}. Transitions take place not only between discrete energy levels but also involve transitions between continuous states and discrete levels. Rayleigh transitions between discrete terms are described by the first term in Equation \eqref{eq:ray1}, ($\mathcal{D}_{Ray}$). Rayleigh transitions between discrete levels and continuous states are described by the second term ($\mathcal{C}_{Ray}$). The corresponding Raman terms are given in Equation \eqref{eq:ram1}.}
\label{fig:levels}}
\end{figure*}

In the following discussion, the physical picture is that of an incoming photon of frequency $\nu$, scattered by a single electron system (i.e. a neutral hydrogen or deuterium atom) that is initially in the ground state (Figures \ref{fig:scattering}, \ref{fig:levels}).  Immediately after the scattering event, the hydrogen atom is left in some new state that we call the ``final state'', which correspond to Panels (b) and (d) in Figure~\ref{fig:scattering}. We use the term ``final'' in the text and throughout the mathematics because further processes involving emission, e.g. Panel (e), are of no concern in the following Kramers-Heisenberg calculation. The scattered photon ($\gamma^\prime$ in Figure~\ref{fig:scattering}) has a frequency  $\nu^\prime$.

Our starting point is the Kramers-Heisenberg scattering cross-section, which can be found in e.g. \cite{Sakurai1967, Berestetskii1971}, and is
\begin{eqnarray}
\frac{d \sigma}{d \Omega} = r_0^2 \left( \frac{\nu^{\prime}}{\nu} \right) \left| \ \delta_{if} \mathbf{e}^{(\alpha)} \cdot \mathbf{e}^{(\alpha^\prime)} + \sum_n  \frac{ 2 \pi m_e \nu_{nf} \nu_{ni}}{\hbar} \left[ \frac{(\mathbf{x} \cdot \mathbf{e}^{(\alpha^\prime)})_{fn} (\mathbf{x} \cdot \mathbf{e}^{(\alpha)})_{ni}}{\nu_{ni} - \nu + i \Gamma/4 \pi} + \frac{(\mathbf{x} \cdot \mathbf{e}^{(\alpha)})_{fn} (\mathbf{x} \cdot \mathbf{e}^{(\alpha^\prime)})_{ni}}{\nu_{ni} + \nu^{\prime}} \right] \  \right|^2 \,.
\label{eq:cross-section}
\end{eqnarray}
In \eqref{eq:cross-section}, $f, n, i$ refer to the final, intermediate, and initial states of the atom, $r_0$ is the classical electron radius, $\nu$ is the frequency of the incoming photon whose direction is given by the unit vector component $\mathbf{e}^{(\alpha)}$ and $\nu^\prime$,  $\mathbf{e}^{(\alpha^\prime)}$ are the corresponding quantities for the outgoing photon. $\delta_{if}$ is the Dirac delta function, $m_e$ is the electron mass, $\Gamma$ is the intrinsic linewidth of the intermediate state, and $(\mathbf{x} \cdot \mathbf{e}^{(\alpha)})_{ni} = \langle n|\mathbf{x} \cdot \mathbf{e}^{(\alpha)}|i \rangle$, and $\mathbf{x}$ is the position operator. It may also be helpful to define what we mean by ``final state''; this means the final state of an atom that forms part of the Rayleigh or Raman scattering processes, illustrated in panels (b) and (d) of Figure \ref{fig:scattering}. The summation in equation \ref{eq:cross-section} is taken over all possible states, and
\begin{eqnarray}
\nu_{ni} = \frac{E_n - E_i}{ 2 \pi \hbar} = \nu_\infty \left( \frac{1}{i^2} - \frac{1}{n^2} \right) \,,
\end{eqnarray}
where $\nu_\infty$ is the Lyman limit frequency. The scattering cross-section must allow for discrete energy levels, i.e. those at energies less than the Lyman limit as well as transitions involving continuum states with energies higher than the Lyman limit.

In the bound state region, we have $E_n = -E_\infty/n^2$, where $E_\infty \sim 13.6~eV$ and $n$ is the principal quantum number. To describe the continuum state, i.e. above the Lyman limit, the energy of the free electron is given by $E_{k^\prime} = \hbar^2 k^2/2m_e$, where we have introduced the positive quantity ${k^\prime}$, which is related to the electron energy by $E_{k^\prime} = E_\infty/k^{\prime 2}$. We thus see that $k^\prime = \sqrt{2m_e E_\infty}/\hbar k$, proportional to the inverse of the wavenumber $k$. 
In addition, it is useful to define $\nu_{k^\prime n} = (E_{k^\prime} - E_n) / 2 \pi \hbar$.
Then, Rayleigh scattering, for which $f = 1$ and which comprises two terms, derived by integrating equation \eqref{eq:cross-section} over the solid angle $\Omega$, can be expressed as \citep{Lee1997}
\smallskip
\begin{eqnarray}
\label{eq:ray1s}
&& \sigma_{Ray}(\nu) = \sigma_{Th} \left| \ \sum_{n=2}^\infty |\langle r\rangle_{np,1s}|^2 \left( \frac{\nu \nu_{n1}}{6 \nu_\infty^2 a_0^2} \frac{m_e}{\mu_e} \right) \left( \frac{\nu_\infty}{ \nu_{n1} - \nu  + i \Gamma/4 \pi} + \frac{\nu_\infty}{ \nu_{n1} + \nu} \right) \right. \nonumber \\
&& \qquad \qquad \left. + \int_0^\infty dk^\prime \left[ |\langle r\rangle_{k^\prime p,1s}|^2 \left( \frac{\nu \nu_{k^\prime 1}}{6 \nu_\infty^2 a_0^2} \frac{m_e}{\mu_e} \right) \left( \frac{\nu_\infty}{ \nu_{k^\prime 1} - \nu } + \frac{\nu_\infty}{ \nu_{k^\prime 1} + \nu } \right) \right] \right|^2 \\
\label{eq:ray1}
&& \qquad \qquad \equiv \sigma_{Th} \left| \ \mathcal{D}_{Ray}(\nu) + \mathcal{C}_{Ray}(\nu) \right|^2 \,,
\end{eqnarray}
where
\begin{eqnarray}
\label{eq:thomson}
\sigma_{Th} = 8 \pi r_0^2 /3
\end{eqnarray}
is the Thomson cross section, the initial state is taken to be $1s$, the intermediate state is $np$, due to the selection rule $\Delta l = \pm 1$, the Lyman limit frequency $\nu_{\infty} = E_\infty/2 \pi \hbar$, the Bohr radius $a_0 = 4 \pi \epsilon_0 \hbar^2/\mu_e e^2$, the reduced mass $\mu_e = m_e m_p/m_e+m_p$, and we have defined $\langle r\rangle_{S_1,S_2} = \langle S_1|r|S_2\rangle$, where $S_1$ and $S_2$ refer to all the possible initial, intermediate, and final states, e.g. $\langle r\rangle_{np, 1s} = \langle np|r|1s \rangle$ and  $\langle r\rangle_{f, k^\prime p} = \langle f|r|k^\prime p \rangle$. The first (discrete) term in Equation \eqref{eq:ray1} thus concerns frequencies from the ground state up to the Lyman limit whilst the second (continuous) term allows for all frequencies beyond the Lyman limit.

Following the same procedure as above, the cross-section for Raman scattering (for which the principal quantum number of the final state $n_f > 1$) is
\smallskip
\begin{eqnarray}
\label{eq:ram1s}
&& \sigma_{Ram}^f(\nu) = \frac{\sigma_{Th} \nu^\prime}{\nu}
\left| \ \sum_{n>n_f} \left[ \langle r\rangle_{f,np} \langle r\rangle_{np,1s} \left( \frac{\nu_{nf} \nu_{n1}}{6 \nu_\infty^2 a_0^2} \frac{m_e}{\mu_e} \right) \left( \frac{\nu_\infty}{\nu_{n1} - \nu + i \Gamma/4 \pi} + \frac{\nu_\infty}{\nu_{n1} + \nu^{\prime}} \right) \right] \right. \\
\label{eq:ram1}
&& \qquad \qquad \left. + \int_0^\infty dk^\prime \left[ \langle r\rangle_{f,k^\prime p} \langle r\rangle_{k^\prime p,1s} \left( \frac{\nu_{k^\prime f} \nu_{k^\prime 1}}{6 \nu_\infty^2 a_0^2} \frac{m_e}{\mu_e} \right) \left( \frac{\nu_\infty}{\nu_{k^\prime 1} - \nu} + \frac{\nu_\infty}{\nu_{k^\prime 1} + \nu^{\prime}} \right) \right] \ \right|^2 \nonumber \\
&& \qquad \qquad \equiv  \frac{\sigma_{Th} \nu^{\prime}}{\nu} \left| \ \mathcal{D}_{Ram}^f(\nu) + \mathcal{C}_{Ram}^f(\nu) \ \right|^2 \,.
\end{eqnarray}
The outgoing photon frequency $\nu^\prime = \nu - \nu_{n_f 1}$, where $n_f$ denotes the final state of the atom after scattering. The Raman scattering cross-section is zero when the incoming photon energy is less than the energy gap between $n=1$ and $n_f$, i.e., $\sigma_{Ram}^f = 0$ for $\nu \leq \nu_{n_f 1}$. Thus, the total cross-section is
\begin{eqnarray}
\label{eq:cs_sum}
\sigma_{KH}(\nu) = \sigma_{Ray}(\nu) + \sum_{f} \sigma_{Ram}^f (\nu)
\end{eqnarray}
where the summation is taken over all possible final states satisfying $\nu > \nu_{n_f 1}$ and $n_f \geq 2$.
Note that the final state can be either $f = n_f s$ or $n_f d$ due to the selection rule $\Delta l = \pm 1$.

\subsection{Calculating the transition coefficients for the discrete terms}

Using the hydrogen atom wave function, we obtain
\begin{eqnarray}
\label{eq:coef_rams}
&& \langle r\rangle_{np,n_fs} = - i \int_0^\infty R_{np} R_{n_f s} r^3 dr = 4 i \times (-1)^{n_f} (n n_f)^{\frac{5}{2}} \sqrt{n^2 -1} \frac{(n-n_f)^{n+n_f-4}}{(n+n_f)^{n+n_f}}   \nonumber \\
&& \qquad \qquad \times \left[ F\left(2-n,1-n_f;2;- \frac{4n n_f}{(n-n_f)^2}\right) - \left( \frac{n-n_f}{n+n_f} \right)^2 F\left(-n,1-n_f;2;- \frac{4n n_f}{(n-n_f)^2}\right) \right] a_0 \,, \end{eqnarray}
and
\begin{eqnarray}
\label{eq:coef_ramd}
&& \langle r\rangle_{np,n_fd}  = \sqrt{2} i \int_0^\infty R_{np} R_{n_f d} r^3 dr = i \times (-1)^{n-1} \frac{(2 n n_f)^{\frac{7}{2}} \sqrt{(n^2 -1) (n_f^2 -1) (n_f^2 -4)}}{9} \frac{(n-n_f)^{n+n_f-6}}{ (n+n_f)^{n+n_f}} \nonumber \\
&& \qquad \qquad \times \left[ F\left(3-n_f,2-n;4;- \frac{4n n_f}{(n-n_f)^2}\right) - \left( \frac{n-n_f}{n+n_f} \right)^2 F\left(1-n_f,2-n;4;- \frac{4n n_f}{(n-n_f)^2}\right) \right] a_0  \,,
\end{eqnarray}
where $R_{nl}$ is the radial wave function with the principal quantum number $n$ and the angular momentum quantum number $l$ (see for example equations 52.2 in \cite{Berestetskii1971} and equation \eqref{eq:wf_disc} below), and
\begin{eqnarray}
F(x_1,x_2;x_3;z) = \sum_{j=0}^\infty \frac{(x_1)_j (x_2)_j}{(x_3)_j} \frac{z^j}{j!}
\end{eqnarray}
is the hypergeometric function and $(x_1)_j = x_1 (x_1+1) (x_1+2)...(x_1+j-1)$ is the rising Pochhammer symbol.
For example, we can obtain the transition coefficient for the Rayleigh scattering from equation \eqref{eq:coef_ray} with $n_f =1$, which is
\begin{eqnarray}
\label{eq:coef_ray}
|\langle r\rangle_{np,1s}|^2 = 2^8 n^7 \frac{(n-1)^{2n-5}}{(n+1)^{2n+5}} a_0^2.
\end{eqnarray}

Evaluating equations \eqref{eq:coef_rams}, \eqref{eq:coef_ramd} and then substitution into equations \eqref{eq:ray1} or \eqref{eq:ram1} allows us to calculate the discrete terms. We now need to calculate the continuous terms.

\subsection{Calculating the transition coefficients for the continuous terms}
\label{sec:continuous}

The radial wave function of the hydrogen for the discrete $R_{nl}$ and the continuum state $R_{k^\prime l}$ can be found in \cite{Landau1981Quantum, Bethe1957},
\begin{eqnarray}
\label{eq:wf_disc}
&& R_{nl} = \frac{2}{n^2 (2l+1)!} \sqrt{\frac{(n+l)!}{(n-l-l)!}} \left( \frac{2r}{n} \right)^l e^{-\frac{r}{n}} F \left(-n+l+1, 2l+2, 2r/n \right) \nonumber \\
&& \qquad = \frac{2}{n^2} \sqrt{(n+l)! (n-l-1)!} e^{-\frac{r}{n}} \sum_{j=0}^{n-l-1} \frac{(-1)^j \left( \frac{2r}{n} \right)^{j+l}}{(n-l-1-j)! (2l+1+j)! j!} \,,
\\
\label{eq:wf_cont}
&& R_{k^\prime l}(r) = \frac{2}{\left( k^{\prime}\right)^{\frac{3}{2}} \sqrt{1- e^{-2 \pi k^\prime}}} \prod_{j=1}^l \sqrt{j^2 + \left( k^{\prime}\right)^{ 2}} \frac{\left( -2r / k^\prime \right)^{-l-1}}{2 \pi} \oint e^{2 i r t / k^\prime} \left( t + \frac{1}{2} \right)^{i k^\prime -l -1} \left( t - \frac{1}{2} \right)^{- i k^\prime -l -1} dt \,,
\end{eqnarray}
where
\begin{eqnarray}
F(a,b,z) = \sum_{j=0}^\infty \frac{(a)_j}{(b)_j} \frac{z^j}{j!}
\end{eqnarray}
is the confluent hypergeometric function. Then, we can derive the transition coefficients required in equations \eqref{eq:ray1} and \eqref{eq:ram1} from equations \eqref{eq:wf_disc} and \eqref{eq:wf_cont} using
\begin{eqnarray}
\label{eq:int_conts}
&& \langle r\rangle_{k^\prime p,n_fs} = - i \int_0^\infty R_{k^\prime p} R_{n_f s} r^3 dr  = \frac{-2 i \sqrt{n_f! (n_f-1)!}}{n_f^2} \sum_{j=0}^{n_f-1} \frac{(-1)^j \left( \frac{2}{n_f} \right)^{j}}{(n_f-j-1)! (j+1)! j!} \int_0^\infty R_{k^\prime p} r^{j+3} e^{-\frac{r}{n_f}} dr \\
\label{eq:int_contd}
&& \langle r\rangle_{k^\prime p,n_fd} = \sqrt{2} i \int_0^\infty R_{k^\prime p} R_{n_f s} r^3 dr  = \frac{2 i \sqrt{2(n_f+2)! (n_f-3)!}}{n_f^2}  \sum_{j=0}^{n_f-3} \frac{(-1)^j \left( \frac{2}{n_f} \right)^{j+2}}{(n_f-j-3)! (j+5)! j!} \int_0^\infty R_{k^\prime p} r^{j+5} e^{-\frac{r}{n_f}} dr
\end{eqnarray}
with
\begin{eqnarray}
\label{eq:wf_int}
\int_0^\infty \left[ R_{k^\prime p} r^m e^{-r/n} \right] r^2 dr =  (-1)^m \times 4 \left( k^{\prime}\right)^{ m+ \frac{3}{2}} \sqrt{\frac{\left( k^{\prime}\right)^{ 2}+1}{1- e^{-2 \pi k^\prime}}} \frac{d^m}{d \xi^m} \left[ \frac{e^{-2 k^\prime \arccot (\xi)}}{(\xi^2 +1)^2} \right]_{\xi = \frac{k^\prime}{n}} \,,
\end{eqnarray}
where $m=j+1$, $j+3$ in equations \eqref{eq:int_conts}, and \eqref{eq:int_contd}, respectively, and $\frac{d^m}{d \xi^m}$ denotes the $m^{th}$ derivative of the term in square brackets. The evaluation of equation \eqref{eq:wf_int} for $m=1$ can be found in the textbook \cite{Bethe1957}. 

\begin{figure*}
\centering
\includegraphics[width=0.99\linewidth]{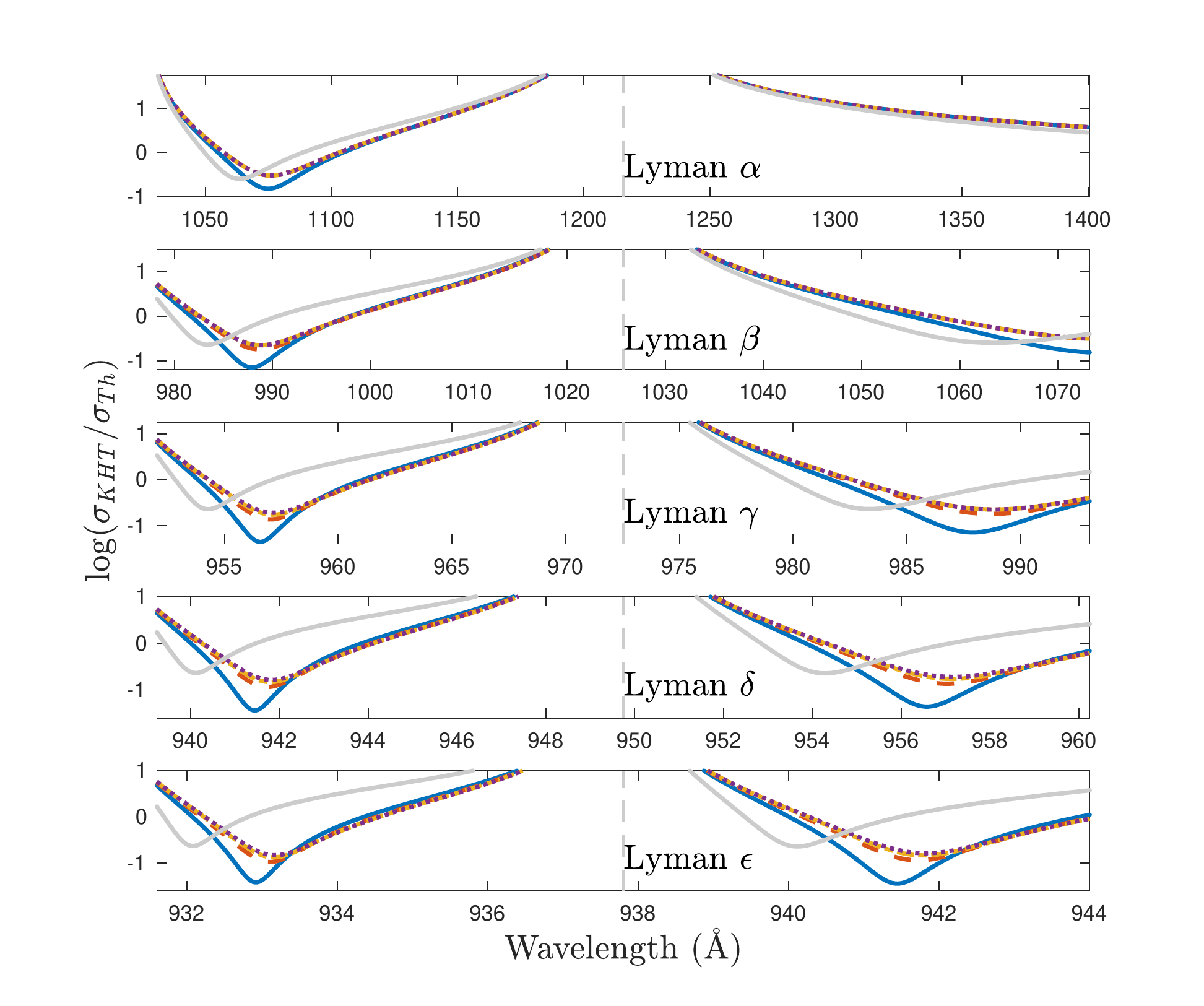}
\caption[caption]{. Illustration of the Kramers-Heisenberg-thermal (KHT) absorption cross-section $\sigma_{KHT}$ (relative to Thomson scattering) for the first 5 lines in the Lyman series. The relevant KHT cross-sections are \eqref{eq:sigma_tot1} and \eqref{eq:sigma_tot2} and the Thomson cross-section is given in equation \eqref{eq:thomson}. The purposes are (i) to illustrate that the continuous terms are of decreasing importance as more continuous terms are added, and (ii) that the same is true down the Lyman series. Hence, we are justified in simplifying (i.e. truncating) the series calculation for continuous terms, as described in Section \ref{sec:cross}. To more clearly distinguish between the various curves, the vertical scale cuts off the line centre (shown in later figures). 5 curves are illustrated in each panel, in which all discrete terms are included, but the number of continuous terms is increased from zero to four:\\\hspace{\textwidth}
Grey solid line: no continuous terms are included. $\sigma(\nu)$ is evaluated from equation~\eqref{eq:cs_sum} with $\mathcal{C}_{Ray}(\nu) = \mathcal{C}_{Ram}^f (\nu) =0$. \\\hspace{\textwidth}
Blue solid line: the most important continuous transition (Rayleigh scattering) is included (i.e. from initial state $i=1s$ to all continuous states and back to the final state, $f=1s$) such that $\sigma(\nu)$ is evaluated from equation~\eqref{eq:cs_sum} with $\sigma_{Ray} = \sigma_{Th} \Big| \mathcal{D}_{Ray}(\nu) + \mathcal{C}_{Ray}(\nu) \Big|^2$ and keeping $\mathcal{C}_{Ram}^f (\nu) =0$. \\\hspace{\textwidth}
Red dashed line: the first two continuous transitions are included (from initial state $i=1s$ to all continuous states, back to $f=1s$ and $f=2s$) such that $\sigma(\nu)$ is evaluated with $\sigma_{Ray} = \sigma_{Th} \Big| \mathcal{D}_{Ray}(\nu) + \mathcal{C}_{Ray}(\nu) \Big|^2$ and $\sigma_{Ram}^f = \frac{\sigma_{Th} \nu^{\prime}}{\nu} \Big| \mathcal{D}_{Ram}^f(\nu) + \mathcal{C}_{Ram}^f (\nu) \Big| $, where $\mathcal{C}_{Ram}^{2s} (\nu)$ is calculated from equation~\eqref{eq:cont} and the remaining $\mathcal{C}_{Ram}^f (\nu) = 0$ for all possible states above $2s$.\\\hspace{\textwidth}
Yellow dot-dashed line: the first three continuous transitions are included (from initial state $i=1s$ to all continuous states, back to $f=1s, 2s$ and $3s$), i.e. the same as the yellow dashed line with non-zero $\mathcal{C}_{Ram}^{3s} (\nu)$.\\\hspace{\textwidth}
Purple dotted line: the first four continuous transitions are included (i.e. from initial state $i=1s$ to all continuous states, back to $f=1s, 2s$, $3s$ and $4s$), i.e. the same as the purple dot-dashed line with non-zero $\mathcal{C}_{Ram}^{4s} (\nu)$.\\\hspace{\textwidth}
The 5 curves illustrated in each panel show that whilst significant changes are made to the cross-section when adding the first continuous term, improvements rapidly diminish with increasing terms -- including the 4$^{th}$ continuous transition makes very little change.}
\label{fig:cont_sigma}
\end{figure*}

Evaluating the continuous term in equation \eqref{eq:ram1} evidently requires evaluating equations \eqref{eq:int_conts} and \eqref{eq:int_contd}. The latter involve summations over $j$ up to $(n_f-l-1)$. Since in principle $n_f \rightarrow \infty$, practical computation forces us to assess the number of summation terms required in order to reach a sufficiently accurate convergence. The transition amplitudes are given by equations \eqref{eq:ray1} and \eqref{eq:ram1} and the overall continuum term is the sum of the Rayleigh and Raman continuum states, $\mathcal{C}_{Ray}(\nu)$ and $\mathcal{C}_{Ram}^f(\nu)$, which are obtained from
\begin{eqnarray}
\label{eq:cont}
\mathcal{C}_{Ray}(\nu) = \int_0^\infty \mathcal{C}_{Ray}(\nu,k^\prime) dk^\prime \qquad \mathrm{and} \qquad \mathcal{C}_{Ram}^f(\nu) = \int_0^\infty \mathcal{C}_{Ram}^f(\nu,k^\prime) dk^\prime \,,
\end{eqnarray}
where 
\begin{eqnarray}
&& \mathcal{C}_{Ray}(\nu,k^\prime) = \int_0^\infty dk^\prime \left| \ \left| \ \langle r\rangle_{k^\prime p,1s} \ \right|^2 \left( \frac{\nu \nu_{k^\prime 1}}{6 \nu_\infty^2 a_0^2} \frac{m_e}{\mu_e} \right) \left( \frac{\nu_\infty}{ \nu_{k^\prime 1} - \nu } + \frac{\nu_\infty}{ \nu_{k^\prime 1} + \nu } \right) \ \right|^2 \,, \\
&& \mathcal{C}_{Ram}^f(\nu,k^\prime) = \int_0^\infty dk^\prime \left| \ \langle r\rangle_{f,k^\prime p} \langle r\rangle_{k^\prime p,1s} \left( \frac{\nu_{k^\prime f} \nu_{k^\prime 1}}{6 \nu_\infty^2 a_0^2} \frac{m_e}{\mu_e} \right) \left( \frac{\nu_\infty}{\nu_{k^\prime 1} - \nu} + \frac{\nu_\infty}{\nu_{k^\prime 1} + \nu^{\prime}} \right) \ \right|^2 \,,
\end{eqnarray}
as shown in equations~\eqref{eq:ray1} and \eqref{eq:ram1}.
Figure \ref{fig:cont_sigma} illustrates the rapid cross-section convergence by the 4th term in $\mathcal{C}_{Ray} (\nu)$ and $\mathcal{C}_{Ram}^f (\nu)$ with $f= 2s$, $3s$ and $4s$ i.e. the green dotted curve is the one adopted.

\section{Modifying the Kramers-Heisenberg model to include Thermal Broadening -- the KHT profile}
\label{sec:broadening}

As far as we are aware, previous investigations of the Kramers-Heisenberg profile have assumed $b=0$, i.e. no account has been taken of thermal broadening.  This is reasonable under many circumstances because the ``Kramers-Heisenberg with thermal broadening'' (hereafter referred to as KHT) and Voigt profiles only begin to differ at very high column densities where line damping effects dominate over thermal broadening effects. Nevertheless, in some applications it is important to include thermal broadening. When calculating the entire Lyman series, line strengths decline rapidly for higher order lines, such that thermal broadening becomes increasingly important. When simultaneously modelling many HI lines in the Lyman series in order to extract best-fit parameters from observational data, ignoring thermal broadening, even in high column density systems, will produce incorrect results.

\begin{figure*}
\centering
\floatbox[{\capbeside\thisfloatsetup{capbesideposition={right,center},capbesidewidth=-4 cm}}]{figure}[\FBwidth]
{\includegraphics[width=8.5cm]{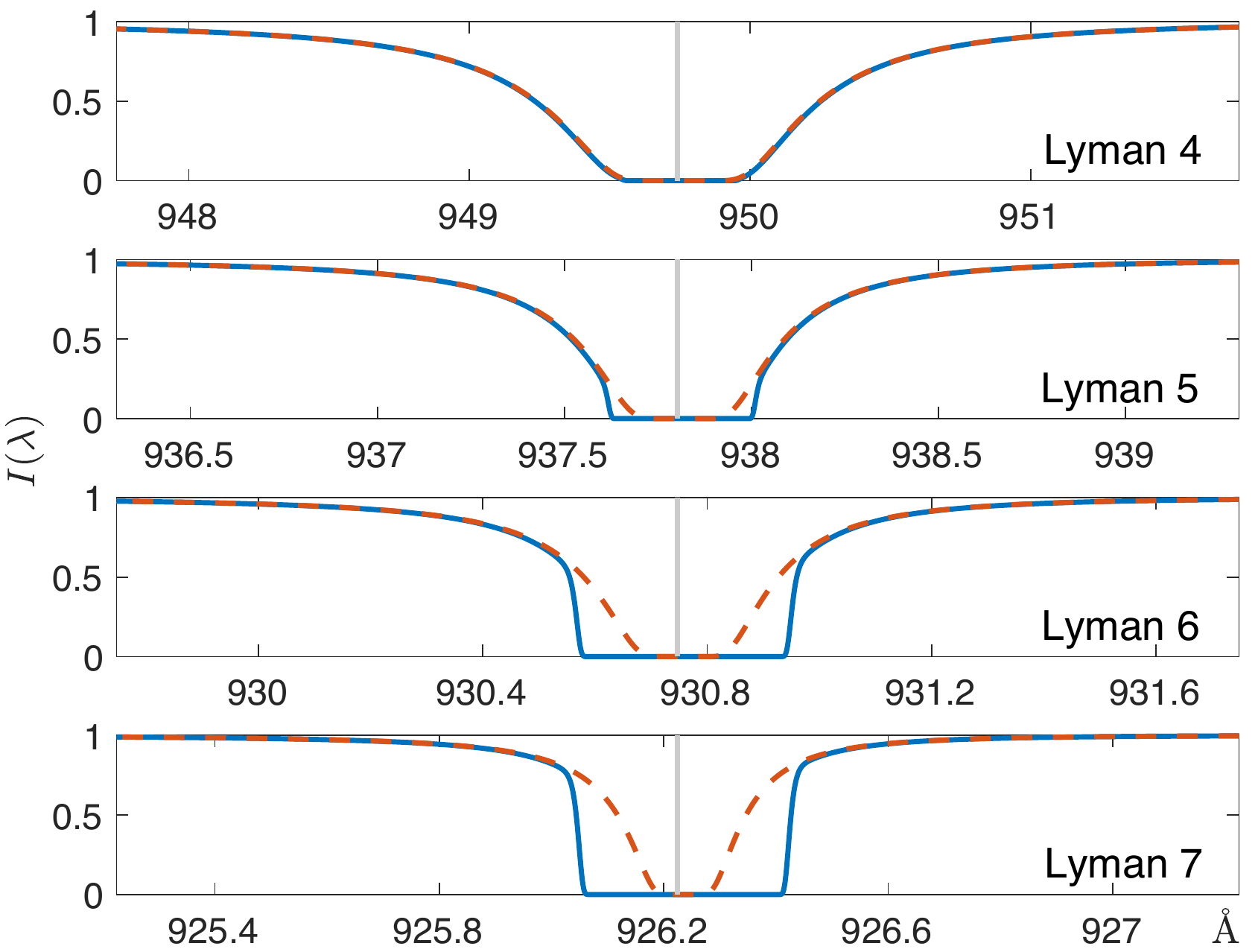}}
{\caption{. Lyman series profiles with neutral hydrogen column density $5 \times 10^{21}$ atoms/cm$^2$, illustrating the intensity as a function of wavelength with $b=16$ km/s (blue solid line) and $b=0$ km/s  (red dashed line). Whilst the difference is negligible at Lyman-$\alpha$ to $\gamma$ for high column density absorption systems, the rapid weakening of higher order Lyman lines means that the inclusion of thermal broadening becomes increasingly important (Section \ref{sec:broadening}).}
\label{fig:broadening}}
\end{figure*}

There are important implications when modelling neutral Lyman lines from DI and HI in order to derive the primordial deuterium abundance in quasar spectra (e.g. \cite{Zavarygin2018}). Since the DI column density is typically of order $10^{5}$ times smaller than HI, thermal broadening becomes important even at Lyman-$\alpha$ for DI and must be accounted for. As an example, we found that for an absorption system with $\log N(HI) = 21.7$, the Lyman profiles with and without thermal broadening (comparing for illustrative purposes $b=0$ with $b=15$), the Lyman lines begin to differ markedly at Lyman 5 (see Figure \ref{fig:broadening}).

Finally, it is of interest to illustrate how the convolution of Kramers-Heisenberg and Maxwellian profiles reduce to the Voigt model at appropriate parameter limits, as shown in Equations~\eqref{eq:ray_voigt1} and \eqref{eq:ram_voigt1}. For all the reasons above, we now modify the Kramers-Heisenberg profile to include thermal broadening.

\subsection{Thermal broadening: Voigt profile}

The Maxwellian distribution expresses the distribution of particles of mass $m$, in a gas in local thermodynamic equilibrium, with velocities in the $z-$direction in the interval $[v_z,\ v_z + dv_z]$,
\begin{eqnarray}
\label{eq:Boltz}
p(v_z) = \frac{e^{-v_z^2/b^2}}{\sqrt{\pi} b} \,,
\end{eqnarray}
where $b= \sqrt{2kT/m}$. For each particle, the observed Doppler shifted frequency $\nu_0^\prime$ is related to the rest-frame frequency of the atom $\nu_0$ by $\nu_0^\prime = \nu_0 \sqrt{(1+v_z)/(1-v_z)} \simeq \nu_0 (1 + v_z/c)$.
The absorption cross-section, commonly used in astrophysics, is the convolution of the Lorentzian~\eqref{eq:lorz} and the Maxwellian distribution~\eqref{eq:Boltz}, which is
\begin{eqnarray}
\label{eq:voigt_sigma}
&& \sigma_{V} = \sqrt{\frac{3 \pi \sigma_{Th}}{8}} c f \int_{-\infty}^\infty dv_z \left[ \frac{p(v_z) \Gamma/4\pi^2}{\left(\nu-\nu_0^\prime \right)^2 + \left(\Gamma/ 4\pi \right)^2} \right] \nonumber \\
&& \qquad = \sqrt{\frac{3 \sigma_{Th}}{8}} \frac{c f}{\Delta \nu_0} \times \frac{a}{\pi} \int_{-\infty}^\infty dy \left[ \frac{e^{-y^2}}{( u - y )^2 + a^2} \right] = \sqrt{\frac{3 \sigma_{Th}}{8}} \frac{c f}{\Delta \nu_0} H(a,u) \,,
\end{eqnarray}
where $\nu$ is the frequency of the incoming photon with respect to the ensemble of atoms and $H(a,u)$ is the familiar Voigt function, with
\begin{eqnarray}
a= \frac{\Gamma}{4 \pi \Delta \nu_0} \,, \quad u = \frac{\nu-\nu_0}{ \Delta \nu_0} \quad \mathrm{and} \quad \Delta \nu_0 = \frac{b \nu_0}{c} \,.
\end{eqnarray}

\subsection{Thermal broadening: KHT for Rayleigh Scattering}
\label{sec:ray}

Here we extend the work of \cite{lee2003asymmetric} to include the effects of thermal broadening in the Kramers-Heisenberg profile. The Rayleigh component of the absorption cross-section, including thermal broadening, is the convolution of equations \eqref{eq:ray1} and \eqref{eq:Boltz}, and is given by
\begin{eqnarray}
\sigma_{Ray}(\nu) = \sigma_{Th} \int_{-\infty}^\infty d v_z p(v_z) \left| \mathcal{D}_{Ray} + \mathcal{C}_{Ray} \right|^2 \,.
\end{eqnarray}
We now adapt Equation~\eqref{eq:ray1s} to include the effect of atomic (i.e. thermal) motions, $\nu_{n1} \rightarrow \nu_{n1} (1+v_z/c)$ and $\nu_{k^\prime 1} \rightarrow \nu_{k^\prime 1} (1+v_z/c)$, and change the variable $v_z = by$, to obtain
\begin{eqnarray}
\label{eq:ray5}
\sigma_{Ray}(\nu) = \sigma_{Th} \int_{-\infty}^{\infty} dy \frac{e^{-y^2}}{\sqrt{\pi}} \ \left| \ \sum_{n=2}^\infty A_n^{\mathcal{D}} \left[ \frac{\nu_{\infty}}{ \nu_{n1} \left(1+\frac{by}{c} \right) - \nu + \frac{i \Gamma}{4 \pi}} + \frac{\nu_{\infty}}{ \nu_{n1} \left(1+\frac{by}{c} \right) + \nu} \right] \right. \nonumber \\
\left. + \int_{0}^{\infty} dk^\prime \left[ A_{k^\prime}^{\mathcal{C}} \left( \frac{\nu_\infty}{ \nu_{k^\prime 1} \left(1+\frac{by}{c} \right) - \nu } + \frac{\nu_\infty}{ \nu_{k^\prime 1} \left(1+\frac{by}{c} \right) + \nu } \right) \right] \ \right|^2 \,,
\end{eqnarray}
where
\begin{eqnarray}
\label{eq:coef_a}
A_n^{\mathcal{D}} = \frac{|\langle r\rangle_{np,1s}|^2}{6 a_0^2} \frac{\nu_{n1}}{ \nu_{\infty}} \frac{\nu}{ \nu_{\infty}} \frac{m_e}{\mu_e} \,, \qquad
A_{k^\prime}^{\mathcal{C}} = \frac{|\langle r\rangle_{k^\prime p,1s}|^2}{6 a_0^2} \frac{ \nu_{k^\prime 1}}{\nu_\infty} \frac{\nu}{ \nu_{\infty}} \frac{m_e}{\mu_e} \,.
\end{eqnarray}

\subsection{Cross-Section evaluation for Rayleigh scattering}

Practical computation of the KHT cross-section requires numerical approximations, analogous in some ways to the series expansion expressions used when calculating the Voigt function \citep{Harris1948}.

Equation \eqref{eq:ray5} can be simplified using two arguments:
(i) Due to the strong suppression by $e^{-y^2}$, the integration in equation \eqref{eq:ray5} can be effectively written as
\begin{eqnarray}
\label{eq:yc}
\int_{-\infty}^\infty e^{-y^2} \left\{ ... \right\} dy \approx \int_{-y_c}^{y_c} e^{-y^2} \left\{ ... \right\} dy \,,
\end{eqnarray}
where $y_c \lesssim \mathcal{O}(10)$ is the cut-off of the integration. (ii) The atomic velocity is small compared to the speed of light. These two properties permit a Taylor series expansion as a power of $b/c$. However, as with numerical approximation of the Voigt function \citep{Harris1948}, it is necessary to use different treatments near to and far from the line centre.

Consider the denominators in the first terms behind the integrals in equation \eqref{eq:ray5}, 
\begin{eqnarray}
\nu_{n1} \left(1+\frac{by}{c} \right) - \nu + \frac{i \Gamma}{4 \pi} = (\nu_{n1} - \nu) \left(1+\frac{b \nu_{n1}}{c(\nu_{n1} - \nu)} y \right) + \frac{i \Gamma}{4 \pi} \sim (\nu_{n1} - \nu) \left(1 - \frac{y}{u_{_{KH}}} \right) \,,
\end{eqnarray}
where we have used $\Gamma \lesssim 10^9$ $s^{-1} \ll \nu_{n1} \sim 10^{15}$ $s^{-1}$ and $u_{_{KH}}=c(\nu - \nu_{n1})/ b \nu_{n1}$. The Taylor expansion is appropriate in that it converges rapidly when $|u_{_{KH}}^{-1}| \ll 1$. We therefore split the practical calculation into two regimes: far from the line centre, i.e. $|u_{_{KH}}^{-1}| \ll 1$, and $\nu \rightarrow \nu_{n1}$.

\bigskip

\noindent {\bf Case I: Far from the line centre, $|u_{_{KH}}^{-1}| \ll 1$}\smallskip

Evaluating equation \eqref{eq:ray5} analytically is difficult so we now first make a simplifying assumption. Since we are not near the line centre, the $i\Gamma/4\pi$ term in the denominator of the first term of \eqref{eq:ray5} is at least 6 orders of magnitude smaller than the the other terms and can be ignored. We next re-arrange each of the 4 terms in equation \eqref{eq:ray5} and then carry out a Taylor series expansion. Consider the first term inside the summation of \eqref{eq:ray5},
\begin{eqnarray}
\label{eq:taylor_condition}
\frac{\nu_{\infty}}{ \nu_{n1} \left(1+\frac{by}{c} \right) - \nu + \frac{i \Gamma}{4 \pi}} \approx 
\frac{\nu_{\infty}}{ \nu_{n1}-\nu}\frac{1}{ \left(1-\frac{b \nu_{n1}}{c(\nu-\nu_{n1})}y \right)} = \frac{\nu_{\infty}}{ \nu_{n1}-\nu} \sum_{m=1}^\infty \left[ \frac{b \nu_{n1}}{c(\nu-\nu_{n1})}y \right]^m
\end{eqnarray}
where since $| b\nu_{n1}y_c/c(\nu_{n1}-\nu) | \ll 1$ (where $y_c$ is the maximum possible value for $y$ -- see \eqref{eq:yc}), a Taylor series expansion is used in the last step. So far we have only dealt with the first term in \eqref{eq:ray5}, but very similar arguments can be used for second, third and fourth terms. For the second term in \eqref{eq:ray5}, since $| b\nu_{n1}y_c/c(\nu_{n1}+\nu)| < |b\nu_{n1}y_c/c(\nu_{n1}-\nu)| $, the same requirement for an efficient expansion is satisfied. The third and fourth terms in \eqref{eq:ray5} automatically satisfy the Taylor series expansion requirement because
$h\nu_{k^{\prime}1} > h\nu_{n1}$.

Eq.~\eqref{eq:ray5} can now be rewritten as
\begin{eqnarray}
\label{eq:ray_taylor}
\sigma_{Ray} &=& \sigma_{Th} \int_{-\infty}^\infty dy \frac{e^{-y^2}}{\sqrt{\pi}} \left| \ \sum_{m=1}^\infty \sum_{n=2}^\infty A_n^{\mathcal{D}} \left[ \frac{\nu_{\infty}}{\nu_{n1} - \nu} \left( \frac{- \nu_{n1}}{\nu_{n1}-\nu} \frac{by}{c} \right)^{m-1} + \frac{\nu_{\infty}}{\nu_{n1} + \nu} \left( \frac{\nu_{n1}}{\nu_{n1} + \nu} \frac{by}{c} \right)^{m-1} \right] \right. \nonumber \\
&+& \left. \sum_{m=1}^\infty \int_{0}^{\infty} dk^\prime A_{k^\prime}^{\mathcal{C}} \left[ \frac{\nu_\infty}{ \nu_{k^\prime 1} - \nu } \left( \frac{- \nu_{k^\prime 1}}{\nu_{k^\prime 1}-\nu} \frac{by}{c} \right)^{m-1} + \frac{\nu_\infty}{ \nu_{k^\prime 1} + \nu } \left( \frac{\nu_{k^\prime 1}}{\nu_{k^\prime 1} + \nu} \frac{by}{c} \right)^{m-1} \right] \ \right|^2 \quad \nonumber \\
& = & \sigma_{Th} \int_{-\infty}^\infty dy \frac{e^{-y^2}}{\sqrt{\pi}} \left| \ \sum_{m=1}^\infty B_m \left( \frac{by}{c} \right)^{m-1} \ \right|^2
\end{eqnarray}
with
\begin{eqnarray}
\label{eq:ray_b1}
B_m &=& \sum_{n=2}^\infty A_n^{\mathcal{D}} \left(\frac{\nu_{\infty}}{\nu_{n1}} \right) \left[ \left( \frac{1}{\nu/\nu_{n1}+1} \right)^m - \left(\frac{1}{\nu/\nu_{n1}-1} \right)^m \right] \nonumber \\
 &+& \int_{0}^{\infty} dk^\prime A_{k^\prime}^{\mathcal{C}} \left(\frac{\nu_{\infty}}{\nu_{k^\prime 1}} \right) \left[ \left( \frac{1}{ \nu/\nu_{k^\prime 1} +1 }  \right)^{m} - \left( \frac{1}{ \nu/\nu_{k^\prime 1} -1} \right)^{m} \right] \,.
\end{eqnarray}
The Gaussian integral gives
\begin{eqnarray}
\int_{-\infty}^\infty \frac{y^{2n} e^{-y^2}}{\sqrt{\pi}} dy = \frac{(2n-1)!!}{2^n} \quad \mathrm{and} \quad \int_{-\infty}^\infty \frac{y^{2n+1} e^{-y^2}}{\sqrt{\pi}} dy = 0 \,.
\end{eqnarray}
Thus, the cross section in Eq.~\eqref{eq:ray_taylor} can be evaluated
\begin{eqnarray}
\label{eq:ray_taylor2}
\sigma_{Ray} = \sigma_{Th} \left[ B_1^2 + \frac{b^2}{2 c^2} \left( 2 B_1 B_3 + B_2^2 \right) + \frac{3 b^4}{4 c^4} \left( 2 B_1 B_5 + 2 B_2 B_4 + B_3^2 \right) + \cdots \right] \,,
\end{eqnarray}
where the evaluation of the coefficients $B_m$ is discussed in Section \ref{sec:eval_kh}.

A technical point concerning practical calculation for Case 1: some arbitrary frequency far from the centre of one particular Lyman transition can of course fall exactly on the frequency of some different Lyman transition. Whenever $\nu = \nu_{n1}$, for all $n$, the first term in \eqref{eq:ray5} blows up and Taylor series expansion fails. Thus in terms of practical calculation, the expansion method described here for Case 1 applies {\it only} when $\nu$ is not close to any value of $\nu_{n1}$. When $\nu$ does coincide with $\nu_{n1}$, for all $n$, Case II is implemented.

\bigskip

\noindent {\bf Case II: Near the line centre, $\nu \rightarrow \nu_{n1}$}\smallskip

The denominator of the first term in Eq.~\eqref{eq:ray5} approaches zero, so the $i\Gamma/4\pi$ term cannot be ignored and the Taylor series expansion used in Case~I cannot be applied in the same way to the resonance state. We therefore have to split Eq.~\eqref{eq:ray5} into two parts, one being the non-resonance state, the other being the resonance state (the derivation of equation \eqref{eq:ray6} is shown in Appendix \ref{sec:app_case2}),
\begin{eqnarray}
\label{eq:ray6}
\sigma_{Ray}(\nu) = \sigma_{Th} \int_{-\infty}^\infty dy \frac{e^{-y^2}}{\sqrt{\pi}} \left| \ \frac{\nu_\infty}{\Delta \nu_{n1}} \frac{ A_n^{\mathcal{D}} }{ y - u + i a_n} + \sum_{m=1}^\infty B_m^\prime \left( \frac{by}{c} \right)^{m-1} \ \right|^2 \,,
\end{eqnarray}
where $\Gamma(n)$ is derived in equation \eqref{eq:app_gamtot} and
\begin{eqnarray}
u=\frac{\nu - \nu_{n1}}{\Delta \nu_{n1}} \,, \quad \Delta \nu_{n1} = \frac{b \nu_{n1}}{c}  \,, \quad a_n = \frac{\Gamma(n)}{4 \pi \Delta \nu_{n1}} \,,
\end{eqnarray}
and
\begin{eqnarray}
\label{eq:ray_b2}
B_m^\prime = B_m + A_n^{\mathcal{D}} \left(\frac{\nu_\infty}{\nu_{n1}} \right) \left(\frac{1}{\nu/\nu_{n1}-1} \right)^m\,.
\end{eqnarray}
Note that in equation \eqref{eq:ray6}, the first term within the absolute brackets, i.e. the term that is not inside the summation, corresponds physically to the situation where the incoming photon frequency is close to the line centre frequency.  This term by itself represents the usual Voigt function case. We evaluate the square of equation \eqref{eq:ray6} in order to calculate the cross-section,
\begin{eqnarray}
\label{eq:ray_voigt}
\sigma_{Ray}(\nu) = \sigma_{Th} \int_{-\infty}^\infty \frac{e^{-y^2}}{\sqrt{\pi}} dy \left[ (A_n^{\mathcal{D}})^2 \frac{ \left( \nu_\infty / \Delta \nu_{n1} \right)^2}{ (u - y)^2 + a_n^2} + \frac{\nu_\infty}{\Delta \nu_{n1}} \frac{ 2 A_n^{\mathcal{D}} (y-u) }{(y-u)^2 + a_n^2} \sum_{m=1}^\infty B_m^\prime \left( \frac{by}{c} \right)^{m-1} + \sum_{m_1=1}^\infty \sum_{m_2=1}^\infty B_{m_1} B_{m_2} \left( \frac{by}{c} \right)^{m_1+m_2 -2} \right] \,.
\end{eqnarray}
The first term in equation \eqref{eq:ray_voigt} can be expressed as
\begin{eqnarray}
\label{eq:ray_voigt2}
\sigma_{Th} \int_{-\infty}^\infty \frac{e^{-y^2}}{\sqrt{\pi}} dy \left[ (A_n^{\mathcal{D}})^2 \frac{ \left( \nu_\infty / \Delta \nu_{n1} \right)^2}{ (u - y)^2 + a_n^2} \right] = \sigma_{Th} \frac{\pi^{\frac{3}{2}} \nu_{n1}^2 \nu^2 }{\Gamma(n) \Delta \nu_{n1} \nu_\infty^2} \frac{m_e^2}{\mu_e^2} \frac{| \langle r\rangle_{np,1s}|^4}{9 a_0^4} H(a,u) \,,
\end{eqnarray}
where equation \eqref{eq:coef_a} has been used and $H(a,u)$ is the Voigt function. Substituting equations \eqref{eq:app_ray2} and \eqref{eq:app_gam_ray} into \eqref{eq:ray_voigt2}, we obtain the Voigt profile with an additional factor $\Gamma_{Ray}(n)/\Gamma(n)$,
\begin{eqnarray}
\label{eq:ray_voigt3}
\sigma_{Th} \int_{-\infty}^\infty \frac{e^{-y^2}}{\sqrt{\pi}} dy \left[ (A_n^{\mathcal{D}})^2 \frac{ \left( \nu_\infty / \Delta \nu_{n1} \right)^2}{ (u - y)^2 + a_n^2} \right] = \frac{\Gamma_{Ray}(n)}{ \Gamma(n)} \sqrt{\frac{3\sigma_{Th}}{8}} \frac{c f(n) H(a,u)}{\Delta \nu_{n1}} \,.
\end{eqnarray}
In practical evaluation of equation~\eqref{eq:ray_voigt}, the last double summation term is small compared to the other two terms and can be ignored if required. Also, since the summation terms of Equation~\eqref{eq:ray_voigt} are diminishingly small for $m \geq 3$, it is reasonable to include $m=1$ and $2$ only, leading to
\begin{eqnarray}
\label{eq:ray_voigt1}
\sigma_{Ray}(\nu) \simeq \frac{\Gamma_{Ray}(n)}{ \Gamma(n)} \sqrt{\frac{3\sigma_{Th}}{8}} \frac{c f(n) H(a,u)}{\Delta \nu_{n1}} + \sigma_{Th} \frac{2 A_n^{\mathcal{D}} \nu_\infty}{\sqrt{\pi} \Delta \nu_{n1}} \left[ B_1^\prime \int_{-\infty}^\infty \frac{(y-u) e^{-y^2}}{(y-u)^2 + a_n^2} dy + B_2^\prime \int_{-\infty}^\infty \frac{(y-u) e^{-y^2}}{(y-u)^2 +a_n^2} \left( \frac{by}{c} \right) dy \right] \,,
\end{eqnarray}
where $\Gamma_{Ray}(n)$ and $\Gamma(n)$ are defined in Equations~\eqref{eq:app_gam_ray} and \eqref{eq:app_gamtot}.

\subsection{Thermal broadening; KHT for Raman Scattering}

Here we again extend the work of \cite{lee2003asymmetric} to include the effects of thermal broadening but now for Raman scattering. The Raman component of the absorption cross-section, including thermal broadening, is the convolution of equations \eqref{eq:ram1} and \eqref{eq:Boltz}, and is
\begin{eqnarray}
\sigma_{Ram}^f(\nu) = \sigma_{Th} \int_{-\infty}^\infty d v_z \frac{\nu^\prime p(v_z)}{\nu} \left| \mathcal{D}_{Ram}^f + \mathcal{C}_{Ram}^f \right|^2 \,.
\end{eqnarray}
We now adapt Equation~\eqref{eq:ram1s} to include the effect of atomic (i.e. thermal) motions, $\nu_{n1} \rightarrow \nu_{n1} (1+v_z/c)$ and $\nu_{k^\prime 1} \rightarrow \nu_{k^\prime 1} (1+v_z/c)$, and change the variable $v_z = by$, i.e. we have the analogous equation to \eqref{eq:ray5} but now for Raman,
\begin{eqnarray}
\label{eq:ram2}
\sigma_{Ram}^f(\nu) = \sigma_{Th} \int_{-\infty}^{\infty} dy \frac{e^{-y^2}}{\sqrt{\pi}} \frac{\nu^\prime}{\nu} \ \left| \ \sum_{n>n_f} P_n^{\mathcal{D}}(f) \left[ \frac{\nu_{\infty}}{ \nu_{n1} \left(1+\frac{by}{c} \right) - \nu + \frac{i \Gamma}{4 \pi}} + \frac{\nu_{\infty}}{ \nu_{n1} \left(1+\frac{by}{c} \right) + \nu^\prime} \right] \right. \nonumber \\
\left. + \int_{0}^{\infty} dk^\prime \left[ P_{k^\prime}^{\mathcal{C}}(f) \left( \frac{\nu_\infty}{ \nu_{k^\prime 1} \left(1+\frac{by}{c} \right) - \nu } + \frac{\nu_\infty}{ \nu_{k^\prime 1} \left(1+\frac{by}{c} \right) + \nu^\prime } \right) \right] \ \right|^2 \,,
\end{eqnarray}
where $f$ again denotes the final state and can be either $n_f s$ or $n_f d$ and
\begin{eqnarray}
P_n^{\mathcal{D}}(f) = \frac{ \langle r\rangle_{n_f s (d),np} \langle r\rangle_{np,1s}}{6 a_0^2} \frac{\nu_{n1}}{ \nu_{\infty}} \frac{\nu_{nf}}{ \nu_{\infty}} \frac{m_e}{\mu_e} \,, \qquad
P_{k^\prime}^{\mathcal{C}}(f) = \frac{\langle r\rangle_{n_f s (d),k^\prime p} \langle r\rangle_{k^\prime p,1s}}{6 a_0^2} \frac{ \nu_{k^\prime 1}}{\nu_\infty} \frac{\nu_{k^\prime f}}{ \nu_{\infty}} \frac{m_e}{\mu_e} \,.
\end{eqnarray}

\subsection{Cross-Section evaluation for Raman scattering}
Comparing equations \eqref{eq:ray1} and \eqref{eq:ram1}, we see that these two equations are very similar. We can borrow the results in equations \eqref{eq:ray_taylor2} and \eqref{eq:ray_voigt1} accounting for the additional Rayleigh $\rightarrow$ Raman factors already given in \eqref{eq:ray1} $\rightarrow$ \eqref{eq:ram1}, i.e. (i) $(\nu^\prime/\nu)$, (ii) $(\nu/\nu_\infty) \rightarrow \left( \nu_{nf} / \nu_\infty \right)$, (iii) $\sum\limits_{n=2}^{\infty} \rightarrow \sum\limits_{n>n_f}$, and (iv) 
\begin{eqnarray}
\sum_n |\langle r\rangle_{np,1s}|^2 \rightarrow \sum_{n > n_f} \langle r\rangle_{n_f s,np} \langle r\rangle_{np,1s} \quad
\mathrm{or} \quad \sum_n |\langle r\rangle_{np,1s}|^2 \rightarrow \sum_{n > n_f} \langle r\rangle_{n_f d,np} \langle r\rangle_{np,1s} \,,
\end{eqnarray}
and obtain
\begin{eqnarray}
\label{eq:ram_taylor}
\left( \frac{\nu - \nu_{n1}}{\nu} \right) \gg 0 : \quad && \sigma_{Ram}^f = \sigma_{Th} \int_{-\infty}^\infty dy \frac{e^{-y^2}}{\sqrt{\pi}} \frac{\nu^\prime}{\nu} \left| \ \sum_{m=1}^\infty Q_m^f \left( \frac{by}{c} \right)^{m-1} \ \right|^2 \,, \\
\label{eq:ram_voigt1}
\left( \frac{\nu - \nu_{n1}}{\nu} \right) \rightarrow 0 : \quad && \sigma_{Ram}^f \simeq \frac{\Gamma_{Ram}^{s(d)}(n,n_f)}{\Gamma(n)} \sqrt{\frac{3\sigma_{Th}}{8}} \frac{c f_n H(a,u)}{\Delta \nu_{n1}} \nonumber \\
&& \qquad + \sigma_{Th} \left( \frac{\nu^\prime}{\nu} \right) \frac{2 \nu_\infty P_n^{\mathcal{D}}(f)}{\sqrt{\pi} \Delta \nu_{n1}} \left[ Q_1^{\prime f} \int_{-\infty}^\infty \frac{(y-u) e^{-y^2}}{(y-u)^2 + a_n^2} dy + Q_2^{\prime f} \int_{-\infty}^\infty \frac{(y-u) e^{-y^2}}{(y-u)^2 + a_n^2} \left( \frac{by}{c} \right) dy \right] \,, \qquad
\end{eqnarray}
where $\Gamma_{Ram}^{s(d)}(n,n_f)$ is defined in Equations~\eqref{eq:gam_ram1} and \eqref{eq:gam_ram2}, and
\begin{eqnarray}
\label{eq:ram_q1}
Q_m^f &=& \sum_{n>n_f} P_n^{\mathcal{D}}(f) \left(\frac{\nu_{\infty}}{\nu_{n1}} \right) \left[ \left( \frac{1}{\nu^\prime / \nu_{n1}+1} \right)^m - \left(\frac{1}{\nu/\nu_{n1}-1} \right)^m \right] \nonumber \\
 && \qquad + \int_{0}^{\infty} dk^\prime P_{k^\prime}^{\mathcal{C}}(f) \left(\frac{\nu_{\infty}}{\nu_{k^\prime 1}} \right) \left[ \left( \frac{1}{ \nu^\prime / \nu_{k^\prime 1} +1 }  \right)^{m} - \left( \frac{1}{ \nu/\nu_{k^\prime 1} -1} \right)^{m} \right] \,,\\
\label{eq:ram_q2}
Q_m^{\prime f} &=& Q_m^f + P_n^{\mathcal{D}}(f) \left(\frac{\nu_\infty}{\nu_{n1}} \right) \left(\frac{1}{\nu^\prime / \nu_{n1}-1} \right)^m \,.
\end{eqnarray}

\section{Practical calculation of the KHT cross-section}
\label{sec:eval_kh}

\begin{figure*}
\centering
\floatbox[{\capbeside\thisfloatsetup{capbesideposition={right,center},capbesidewidth=-4 cm}}]{figure}[\FBwidth]
{\includegraphics[width=1.0\linewidth]{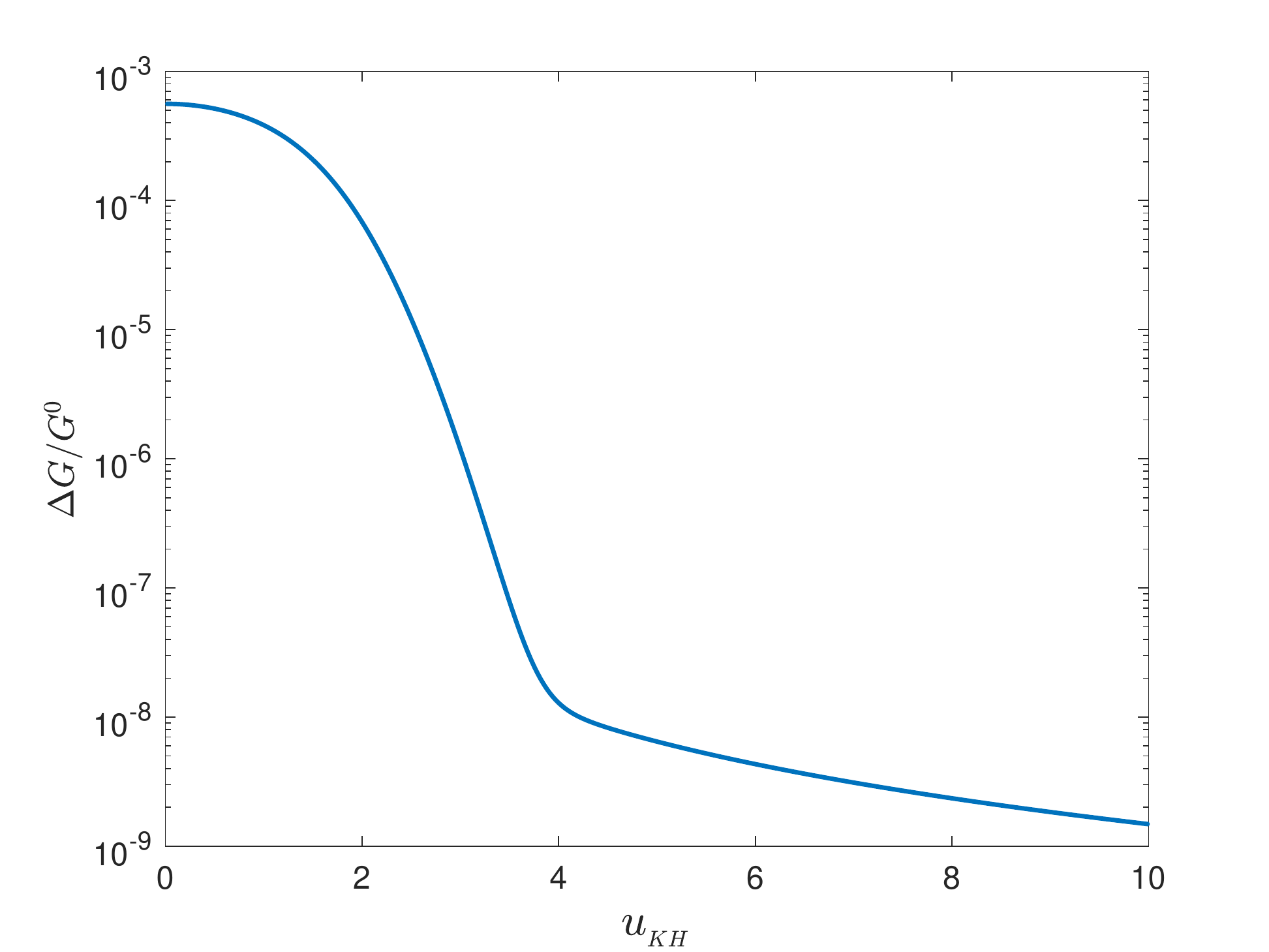}}
{\caption{. Illustration of the fractional error $\Delta G/G^0=(G(u)-G^0(u,a_n))/G^0(u,a_n)$ in computing the function $G^0(u,a_n)$ caused by ignoring $a^2_n$. 
This diagram is important in demonstrating the practicality of computing the KHT profile. The basic point is that it is necessary to remove the dependence of the KHT cross-section on the principal quantum number $a_n$, else practical computation is too hard. The details are provided by equations \eqref{eq:int_tot2} to \eqref{eq:int_tot3} and their associated text. The curve illustrated here is specific to the Lyman-$\alpha$ transition for our assumed absorption line width of $b=16$ km/s, for which $a_2 \simeq 3.8 \times 10^{-4}$, but the same argument will apply generally.}
\label{fig:delG}}
\end{figure*}

We demonstrate some tricks for the numerical computation of the KHT cross-section here. We start from the integral terms in equations \eqref{eq:ray_voigt1} and \eqref{eq:ram_voigt1},
\begin{eqnarray}
\label{eq:int_tot1}
&& G^0(u,a_n) = \int_{-\infty}^\infty \frac{(y-u) e^{-y^2}}{(y-u)^2 + a_n^2} dy \,, \\
\label{eq:int_tot2}
&& K^0(u,b,a_n) = \int_{-\infty}^\infty \frac{(y-u) e^{-y^2}}{(y-u)^2 + a_n^2} \left( \frac{by}{c} \right) dy \,.
\end{eqnarray}
The presence of the $a_n^2$ term is problematic because we would need to compute this integral for all $n$ up to $\infty$, which in turn would make numerical integration impractical. We therefore explore the contribution of this term. Let us remove $a_n$ from equation \eqref{eq:int_tot2} and define a new function $G(u)$,
\begin{eqnarray}
G(u) \equiv \int_{-\infty}^\infty \frac{e^{-y^2}}{y-u} dy \,.
\end{eqnarray}
Figure \ref{fig:delG} shows the difference between $G^0(u,a_n)$ and $G(u)$ with $b=16$ km/s and $n=2$ (Lyman-$\alpha$). The values of the first and second terms in equation \eqref{eq:ray_voigt1} with $b=16$ km/s at $|u_{_{KH}}|= 10^{-2}$, 1, 2, 3 and 10 are
\begin{eqnarray}
\label{eq:approx_value1}
\sigma_{Ray}^{(1)} &\equiv& \sqrt{\frac{3\sigma_{Th}}{8}} \frac{c f(n) H(a,u)}{\Delta \nu_{n1}} \sim 5 \times 10^{-14} \,, \ 2 \times 10^{-14} \,, \ 8 \times 10^{-16} \,, \ 7 \times 10^{-18} \,, \ \mathrm{and} \ 1 \times 10^{-19} \ \mathrm{cm^2} \\
\label{eq:int_tot4}
\sigma_{Ray}^{(2)} &\equiv& \frac{2 \sigma_{Th} A_n^{\mathcal{D}} B_1^\prime \nu_\infty}{\sqrt{\pi} \Delta \nu_{n1}} \int_{-\infty}^\infty \frac{(y-u) e^{-y^2}}{(y-u)^2 + a_n^2} dy \sim 1 \times 10^{-22} \,, \ 6 \times 10^{-21} \,, \ 3 \times 10^{-21} \,, \ 2 \times 10^{-21} \,,  \ \mathrm{and} \ 5 \times 10^{-22} \ \mathrm{cm^2} \,,
\end{eqnarray}
indicating that the KHT cross-section is dominated by the resonance energy level. The numerical error caused by taking $G(u)$ instead of $G^0(u,a_n)$ is
\begin{eqnarray}
\left( \frac{G(u)}{G^0(u,a_n)} -1 \right) \times \frac{\sigma_{Ray}^{(2)}}{\sigma_{Ray}^{(1)}} \sim 8 \times 10^{-5} \times \frac{3 \times 10^{-21}}{8 \times 10^{-16}} = 3 \times 10^{-10} \,,
\label{eq:DGoverG}
\end{eqnarray}
where $u=2$ is taken because the maximum of the numerical error occurs at $|u| \sim 2$. Thus the contributions from the second terms in equations \eqref{eq:ray_voigt1} and \eqref{eq:ram_voigt1} are unimportant. This means we are justified in using $G(u)$ in place of $G^0(u,a_n)$. The same argument is applicable to the integration in equation \eqref{eq:int_tot2}, leading to
\begin{eqnarray}
\label{eq:int_tot3}
K^0(u,b,a_n) = \int_{-\infty}^\infty \frac{(y-u) e^{-y^2}}{(y-u)^2 + a_n^2} \left( \frac{by}{c} \right) dy \approx  K(u,b) \equiv \int_{-\infty}^\infty \frac{e^{-y^2}}{(y-u)} \left( \frac{by}{c} \right) dy =  \left( \sqrt{\pi} + u G(u) \right) \times \left( \frac{b}{c} \right) \,.
\end{eqnarray}
Then, we gain the benefit of removing the principal quantum number dependent parameter, $a_n$, from the numerical integration, i.e., only one single look-up table for $G(u)$ is required for the integration terms in equations \eqref{eq:ray_voigt1} and \eqref{eq:ram_voigt1} with negligible change in accuracy.

The KHT cross-section can now be simplified using $\sigma_{KHT} = \sigma_{Ray} + \sum\limits_f \sigma^f_{Ram}$,
\begin{eqnarray}
\label{eq:sigma_tot1}
\left( \frac{\nu - \nu_{n1}}{\nu} \right) \gg 0 : && \sigma_{KHT} = \sigma_{Th} \left\{ B_1^2 + \frac{b^2}{2 c^2} \left( 2 B_1 B_3 + B_2^2 \right) + \frac{3 b^4}{4 c^4} \left( 2 B_1 B_5 + 2 B_2 B_4 + B_3^2 \right) + \cdots \right. \nonumber \\
&& + \sum_f \left. \left(\frac{\nu^\prime}{\nu}\right) \left[ \left(Q_1^f \right)^2 + \frac{b^2}{2 c^2} \left( 2 Q_1^f Q_3^f + \left(Q_2^f \right)^2 \right) + \frac{3 b^4}{4 c^4} \left( 2 Q_1^f Q_5^f + 2 Q_2^f Q_4^f + \left(Q_3^f \right)^2 \right) + \cdots \right] \right\} \,, \\
\label{eq:sigma_tot2}
\left( \frac{\nu - \nu_{n1}}{\nu} \right) \rightarrow 0 : && \sigma_{KHT} \simeq \sigma_{V} + \frac{ 2 \sigma_{Th} \nu_\infty}{\sqrt{\pi} \Delta \nu_{n1}} \left[ A_n \left( B_1^\prime G(u) + B_2^\prime K(u,b) \right) + \sum_f \left( \frac{\nu^\prime}{\nu} \right) P_n^{\mathcal{D}}(f) \left( Q_1^{\prime f} G(u) + Q_2^{\prime f} H(u,b) \right) \right] \,,
\end{eqnarray}
where $\sigma_V$ is the Voigt cross-section (Equation \eqref{eq:voigt_sigma}). The KHT cross-section, $\sigma_{KHT}$, is computed from equations~\eqref{eq:sigma_tot1} and \eqref{eq:sigma_tot2}, and the infinity sums $\sum\limits_f$ and $\sum\limits_n$ can be overcome by taking a suitable cut-off for the principal quantum number $n_c$. Then, we need $2 \times (n_c-1)$ numerical integrations for the continuous transitions, $\int_{-\infty}^\infty \{ ... \} dk^\prime$, for $B_1$, $Q_1^{1s,~2s,~...~,~n_c s}$ and $Q_1^{3d,~,4d,~...~,~n_c d}$, appearing in equations~\eqref{eq:ray_b1} and \eqref{eq:ram_q1}, and the same number of integrations are needed for $B_2$, $Q_2^{1s,~2s,~...~,~n_c s}$ and $Q_2^{3d,~,4d,~...~,~n_c d}$, etc. However, this suggests a massive calculation with long computing time is required to get the numerical result. In other words, practical application to data analysis would be almost impossible if we include a huge number of continuous transitions. Fortunately, the first four continuous transitions are sufficient. This is illustrated in Figure~\ref{fig:cont_sigma} (and discussed previously in Section \ref{sec:continuous}). 

In addition, neither a singularity nor a dramatic change appears in $\mathcal{C}_{Ray}$ and $\mathcal{C}_{Ram}$ at $\nu < \nu_{\infty}$, so these two functions evolve slowly at wavelengths far from the Lyman limit $\lambda_\infty$. As a result, only $4 \times (2n_p +1)$ look-up tables are required to evaluate the KHT cross-section that use $n_p^{th}$ order perturbation for equations \eqref{eq:ray_taylor} and \eqref{eq:ram_taylor}, i.e.,$\sum\limits_{m=1}^\infty \rightarrow \sum\limits_{m=1}^{n_p+1}$.

In Figure~\ref{fig:sigma_tot}, we illustrate our numerical result for $\sigma_{KHT}$ with $n_p=2$, $n_c=21$ for discrete energy levels, and using the first four terms for continuous transitions. Truncating at $n_c=21$ means we only use contributions to the KHT cross-section from transitions Lyman-$\alpha$ to Lyman-20 for the discrete energy levels in equations \eqref{eq:ray_b1}, \eqref{eq:ray_b2}, \eqref{eq:ram_q1} and \eqref{eq:ram_q2}. In our numerical computation (and in the VPFIT implementation), we use KHT at wavelengths above 915{\AA}. Below 915{\AA} and down to Lyman-31 (912.645{\AA}), the absorption cross-section is dominated by the first term in equation \eqref{eq:sigma_tot2} i.e. the cross-section is accurately described using a Voigt function (see equations \eqref{eq:approx_value1} and \eqref{eq:int_tot4}). Below Lyman-31, VPFIT interpolates to the Lyman limit and uses a simple analytic approximation for the cross-section at wavelengths below the Lyman limit. The truncation at $n_p=2$ is justified next.

In calculating the curves in Figure \ref{fig:sigma_tot}, equation \eqref{eq:sigma_tot1} is used for $|u_{_{KH}}| = |(\nu-\nu_{n1})/\Delta \nu_{n1}| > 10$ and equation \eqref{eq:sigma_tot2} is used for $|u_{_{KH}}| \leq 10$. These same ranges are applied for the computation of KHT inside VPFIT. Finally, 25 look-up tables are used, four for the Voigt part of KHT, one for $G(u)$ and twenty for the continuous transitions. The CPU time for calculating the KHT cross-section is around 80 times longer than for a Voigt profile (also using look-up tables). The use of KHT is only needed for high column density systems. How high depends on which lines in the Lyman series are fitted. This point will be quantified in detail in a forthcoming paper but as a guideline, if fitting more than 4 lines in the Lyman series, it would be advisable to use KHT for column densities above, say, $1 \times 10^{20}$ atoms per square centimetre, for high signal to noise and high resolution DLA spectra.

\subsection{Estimating the KHT cross-section precision}

\noindent {\bf Case I: Far from the line centre, $|u_{_{KH}}| > 10$} \smallskip

\begin{figure*}
\centering
\includegraphics[width=0.75\linewidth]{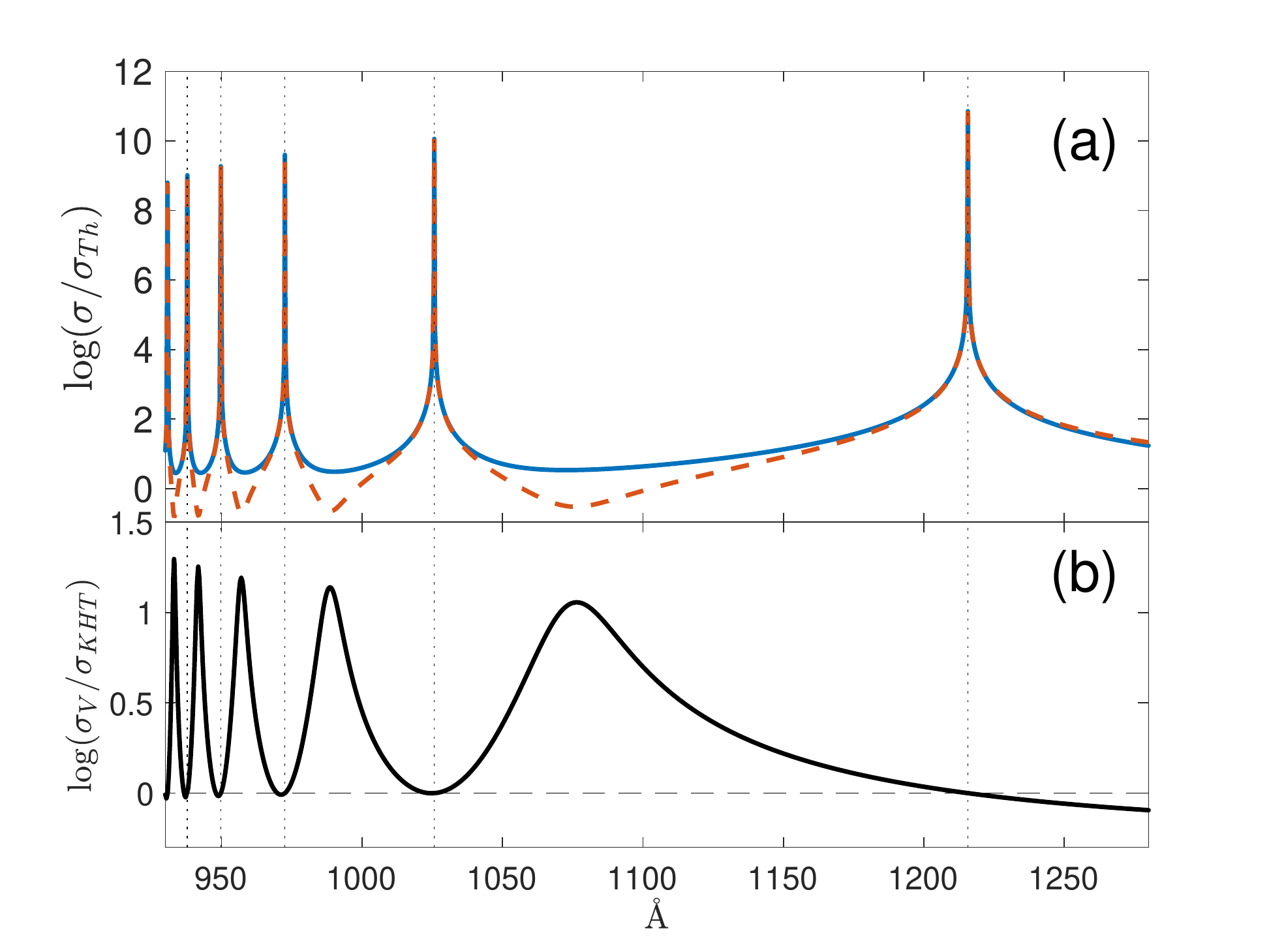}
\\
\includegraphics[width=0.75\linewidth]{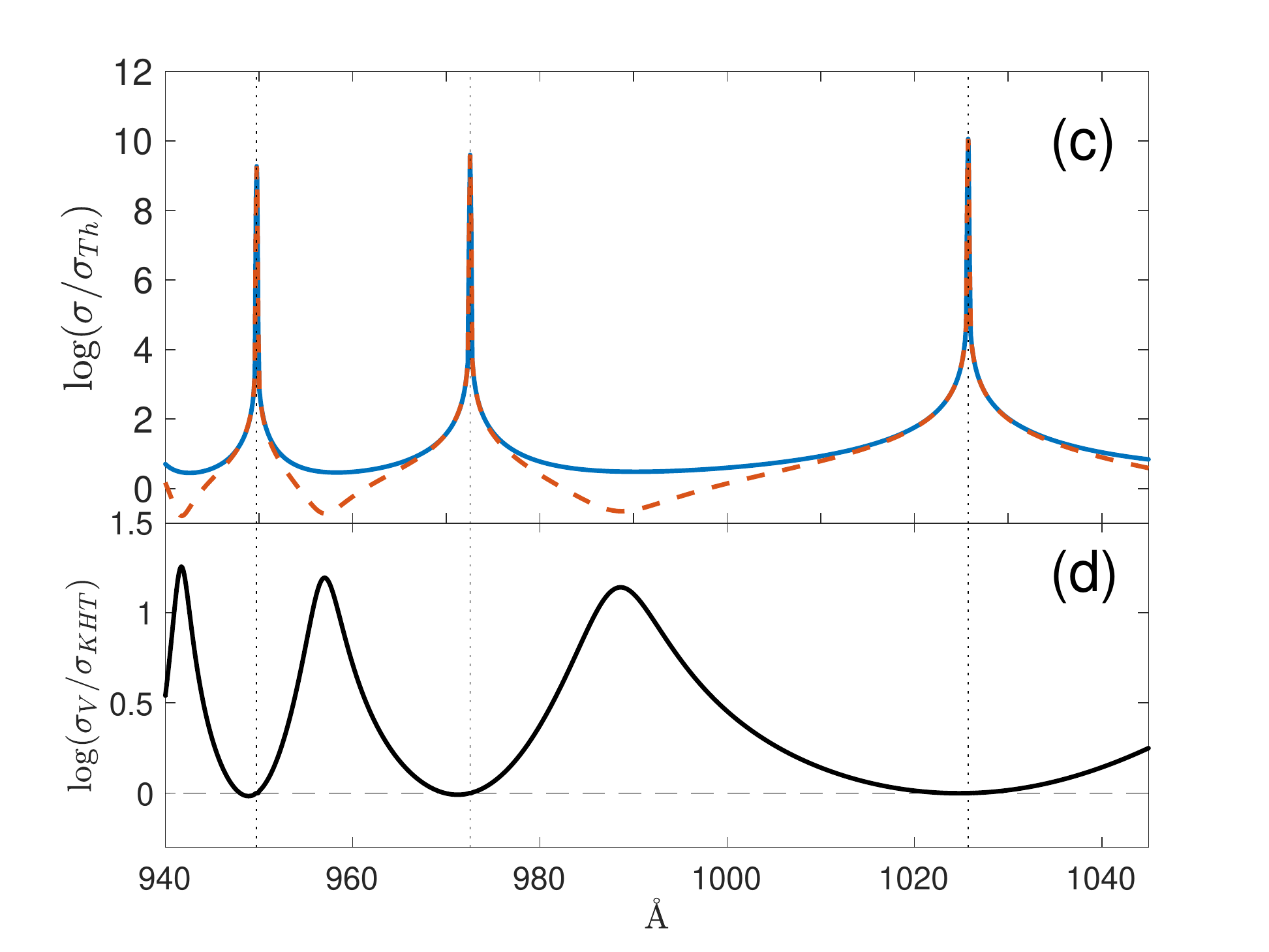}
\caption{. Absorption cross-section for the Lyman series of neutral hydrogen. In panels (a) and (c), the blue continuous line is for a Voigt profile and the red dashed line is KHT. Panel (a) shows Lyman $\alpha, \beta, \gamma, \delta, \epsilon$, and $\zeta$. Panel (c) is a zoom-in on $\beta, \gamma$, and $\delta$. Panels (b) and (d) illustrate the difference between the two cross-sections. Away from the line centre, the KHT absorption cross-section falls below Voigt and there is an asymmetry in the profile wings. This occurs where the amplitude, which is the sum of $\mathcal{D}(\nu)$ and $\mathcal{C}(\nu)$ in equations \eqref{eq:ray1} and \eqref{eq:ram1}, changes sign. The effect has previously been noted by \citet{Lee1997}. The asymmetry changes direction between Lyman-$\alpha$ and the rest of the Lyman series and the degree of asymmetry increases with decreasing wavelength. This effect is cumulative along the Lyman series. In both Voigt and KHT curves, a thermal broadening of 16 km/s is used. The inclusion of thermal broadening is only noticeable very close to the line centre and cannot be clearly seen in this plot so is illustrated in detail in Fig.~\ref{fig:KHthermal}. Between Lyman line centres, we see the KHT cross-section falls to a minimum, exhibiting a different behaviour from the classical Voigt value.}
\label{fig:sigma_tot}
\end{figure*}

We can gain some insight into the numerical precision of the calculated KHT cross-section by inspecting equation \eqref{eq:ray_taylor2}. In particular, we can use this equation to decide on the appropriate number of terms to include in the summation inside the square brackets in \eqref{eq:ray_taylor2}. That is, we can estimate the precision on the $B_m$ coefficients to decide where to truncate the summation of equation \eqref{eq:ray_taylor}, in order to achieve the desired precision on the KHT cross-section calculation. To achieve, we calculate the relative magnitudes of the terms $B_1^2$, $(2B_1 B_3 + B_2^2)$, and so on, in equation \eqref{eq:ray_taylor}. The dominant term when computing $B_m$ is the resonance term, i.e. the term with denominator $\nu/\nu_{n1} -1$ in equation \eqref{eq:ray_b1}. The term with denominator $\nu/\nu_{k\prime 1} -1$ is unimportant because $\nu_{k\prime 1}$ is below the Lyman limit, i.e. is far from $\nu$. The worst Case I precision occurs at the boundary i.e. at $|u_{_{KH}}| \rightarrow 10$, where the value of the denominator is given by
\begin{eqnarray}
u_{_{KH}}=\frac{\nu - \nu_{n1}}{\Delta \nu_{n1}} \quad \Rightarrow \quad \frac{\nu}{\nu_{n1}} -1 \approx \frac{u_{_{KH}}b}{c} \,.
\end{eqnarray}
An order of magnitude value for $B_m$ (from equation~\eqref{eq:ray_b1}) is
\begin{eqnarray}
B_m \sim - A_n^{\mathcal{D}} \left(\frac{\nu_{\infty}}{\nu_{n1}} \right) \left(\frac{1}{\nu/\nu_{n1}-1} \right)^m \approx - A_n^{\mathcal{D}} \left(\frac{\nu_{\infty}}{\nu_{n1}} \right) \left( \frac{c}{u_{_{KH}} b} \right)^m \,.
\end{eqnarray}
Inspecting equation \eqref{eq:ray_taylor2}, the relative values of the terms involving $B$ (i.e. normalising successive terms after the $B_1^2$ by $B_1^2$) are in the approximate ratio (taking $u_{_{KH}} = 10$)
\begin{eqnarray}
\label{eq:accuracy1}
 B_1^2 : \frac{b^2}{2 c^2} \left( 2 B_1 B_3 + B_2^2 \right) : \frac{3 b^4}{4 c^4} \left( 2 B_1 B_5 + 2 B_2 B_4 + B_3^2 \right) : \ \mathrm{etc.} \sim 1:u_{_{KH}}^{-2}:u_{_{KH}}^{-4} : \ \mathrm{etc.} \sim 1: 10^{-2}:10^{-4} : \ \mathrm{etc.} \,,
\end{eqnarray}
since, more generally,
\begin{eqnarray}
\frac{B_{1+m}}{B_1}\left(\frac{b}{c}\right)^m \sim \left( \frac{c}{u_{_{KH}} b} \right)^{m} \left(\frac{b}{c}\right)^m = u_{_{KH}}^{-m} \,.
\end{eqnarray} 
The KHT cross-section in Figure \ref{fig:sigma_tot} has been calculated taking the second order perturbation, giving a relative precision of $\sim 10^{-4}$. Including higher order perturbations would of course improve the accuracy, but the a cost of longer computing time and more look-up tables for continuous transitions.

\bigskip

\noindent {\bf Case II: Near the line centre, $|u_{_{KH}}| \leq 10$} \smallskip

We provide a similar argument to the one used above for $|u_{_{KH}}| > 10$. Equation \eqref{eq:ray_voigt1}  approximates \eqref{eq:ray_voigt} by including terms up to and including $m=2$. By again looking at the relative magnitude of successive terms in the summation, we can approximate the precision on the KHT cross-section. The relative magnitudes of the first 3 terms in equation \eqref{eq:ray_voigt}, i.e. those illustrated in equation \eqref{eq:ray_voigt1}, are (for Lyman-$\alpha$),
\begin{eqnarray}
\label{eq:accuracy2}
1 : 10^{-3} : 10^{-6} : \ \mathrm{etc}.
\end{eqnarray}
where equations \eqref{eq:approx_value1},  \eqref{eq:int_tot4}, and
\eqref{eq:int_tot3} have been used.
For Lyman-10 (as an example), the corresponding numbers are
\begin{eqnarray}
\label{eq:accuracy3}
1 : 1 :  10^{-3} : \ \mathrm{etc}.
\end{eqnarray}

However, as we approach higher order terms, from about Lyman-11 down, the line separation becomes relatively small such that lines blend together and Case I ceases to apply. Also, since line centres remain saturated down to the Lyman limit (at DLA column densities) such that there is zero flux.  Therefore, whilst the KHT precision worsens towards higher order lines, line blending means makes this irrelevant.

\begin{figure*}
\centering
\includegraphics[width=0.8\linewidth]{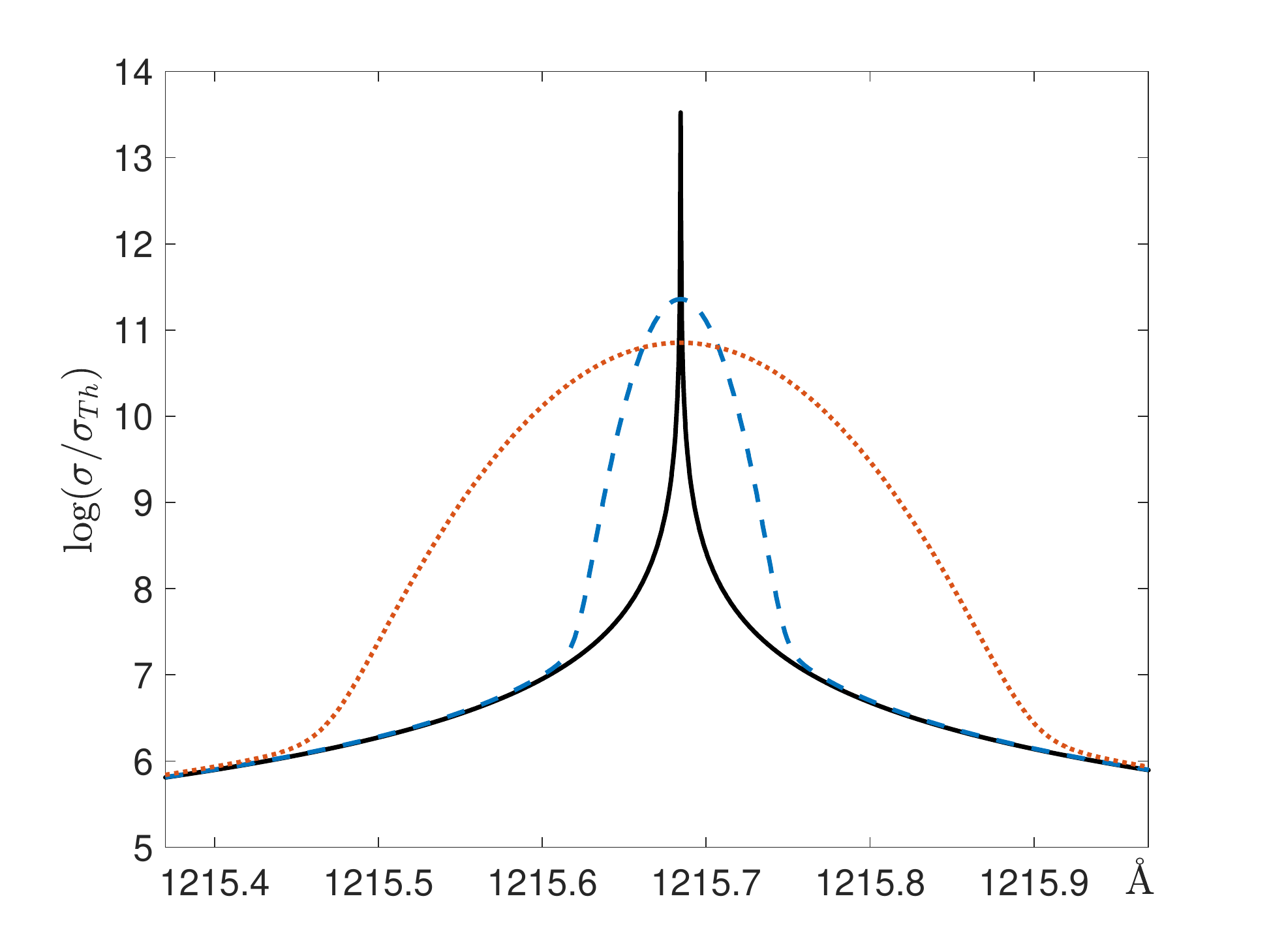}
\caption{. Including thermal broadening in the KHT cross-section. Only Lyman-$\alpha$ is shown. The black continuous curve is without thermal broadening. The blue dashed and red dotted lines include thermal broadening with $b$-parameters of 5 and 16 km/s respectively.}
\label{fig:KHthermal}
\end{figure*}

\section{Discussion}
\label{sec:discussion}

In this paper we have used the Kramers-Heisenberg profile for modelling a single electron atom, i.e. hydrogen or deuterium. The Kramers-Heisenberg profile incorporates natural broadening only and does not allow for thermal broadening. In the same way that the Lorentzian profile was convolved with a Maxwellian to produce the widely-used Voigt profile, we present a new profile which we call the KHT (Kramers-Heisenberg-Thermal) profile. We argue that the KHT model should be used in preference to the Voigt model for high column density absorption systems, particularly when several Lyman series lines are modelled simultaneously. In a separate forthcoming paper we will quantify the systematic errors associated with the Voigt assumption at high column density.

The motivation for doing this work is that the Voigt model is derived from a 2-level oscillator and hence is actually only an approximation, albeit an extremely good one under most circumstances. It nevertheless fails quite severely for high enough atomic column densities, where interactions between multiple discrete levels and with continuum transitions become important, as noted previously (notably by H.-W. Lee and collaborators).

To summarise the main outcomes of this work:

\begin{enumerate}
\item[1.] We have modified the quantum mechanical Kramers-Heisenberg formula to incorporate thermal broadening. Our final equations for practical computation of the Kramers-Heisenberg-thermal (KTH) profile are equations \eqref{eq:sigma_tot1} and \eqref{eq:sigma_tot2}.
\item[2.] Because of the mathematically unwieldy nature of the KHT cross-section, practical calculation requires Taylor series expansions and associated approximations. Analogous to the usual Voigt profile methodologies, we show how using look-up tables can be employed to speed-up computing time.
\item[3.] The required precision for the KHT cross-section can be user-defined, depending on how many terms in equations \eqref{eq:sigma_tot1} and \eqref{eq:sigma_tot2} are included -- examples are given in equations \eqref{eq:accuracy1}, \eqref{eq:accuracy2} and \eqref{eq:accuracy3}. 
\item[4.] Computing speeds are around two orders of magnitude longer than Voigt (using look-up tables in both cases), but this is still easily fast enough for practical applications.
\item[5.] Continuous transitions have been included in our calculations. We include the first 4 continuous terms, showing that these 4 are important but that very little improvement in the cross-section accuracy would be gained by including further terms.
\item[6.] The widely used code, VPFIT, has been updated to include the option to model absorption systems using KHT profiles.
\item[7.] During the course of this work we found that the literature lacked sufficiently accurate damping constants for Lyman series lines beyond Lyman 5. These were therefore calculated and are tabulated in Appendix \ref{app:appdedix2}.
\end{enumerate}

In a forthcoming paper we will illustrate and quantify the systematic biases caused by modelling damped Lyman-$\alpha$ absorption systems using Voigt profiles.

\newpage
\bibliographystyle{mnras}
\bibliography{kh1}

\begin{thebibliography}{}
\makeatletter
\relax
\def\mn@urlcharsother{\let\do\@makeother \do\$\do\&\do\#\do\^\do\_\do\%\do\~}
\def\mn@doi{\begingroup\mn@urlcharsother \@ifnextchar [ {\mn@doi@}
  {\mn@doi@[]}}
\def\mn@doi@[#1]#2{\def\@tempa{#1}\ifx\@tempa\@empty \href
  {http://dx.doi.org/#2} {doi:#2}\else \href {http://dx.doi.org/#2} {#1}\fi
  \endgroup}
\def\mn@eprint#1#2{\mn@eprint@#1:#2::\@nil}
\def\mn@eprint@arXiv#1{\href {http://arxiv.org/abs/#1} {{\tt arXiv:#1}}}
\def\mn@eprint@dblp#1{\href {http://dblp.uni-trier.de/rec/bibtex/#1.xml}
  {dblp:#1}}
\def\mn@eprint@#1:#2:#3:#4\@nil{\def\@tempa {#1}\def\@tempb {#2}\def\@tempc
  {#3}\ifx \@tempc \@empty \let \@tempc \@tempb \let \@tempb \@tempa \fi \ifx
  \@tempb \@empty \def\@tempb {arXiv}\fi \@ifundefined
  {mn@eprint@\@tempb}{\@tempb:\@tempc}{\expandafter \expandafter \csname
  mn@eprint@\@tempb\endcsname \expandafter{\@tempc}}}

\bibitem[\protect\citeauthoryear{{Berestetskii}, {Lifshitz}  \&
  {Pitaevskii}}{{Berestetskii} et~al.}{1971}]{Berestetskii1971}
{Berestetskii} V.~B.,  {Lifshitz} E.~M.,   {Pitaevskii} L.~P.,  1971,
  Relativistic Quantum Theory, First Edition: Volume 4, Part 1, 1 edn

\bibitem[\protect\citeauthoryear{{Bethe} \& {Salpeter}}{{Bethe} \&
  {Salpeter}}{1957}]{Bethe1957}
{Bethe} H.~A.,  {Salpeter} E.~E.,  1957, {Quantum Mechanics of One- and
  Two-Electron Atoms}

\bibitem[\protect\citeauthoryear{Breit}{Breit}{1924}]{Breit1924}
Breit G.,  1924, Nature, 114, 310

\bibitem[\protect\citeauthoryear{Breit}{Breit}{1932}]{Breit1932}
Breit G.,  1932, \mn@doi [Rev. Mod. Phys.] {10.1103/RevModPhys.4.504}, 4, 504

\bibitem[\protect\citeauthoryear{Corney}{Corney}{1977}]{Corney1977}
Corney A.,  1977, Atomic and laser spectroscopy.
Oxford science publications, Clarendon Press, \url
  {https://books.google.co.uk/books?id=JOM0vwEACAAJ}

\bibitem[\protect\citeauthoryear{Dirac}{Dirac}{1927}]{Dirac1927}
Dirac P. A.~M.,  1927, \mn@doi [Proc. Roy. Soc. A]
  {https://doi.org/10.1098/rspa.1927.0071}, 114, 710

\bibitem[\protect\citeauthoryear{{Harris III}}{{Harris III}}{1948}]{Harris1948}
{Harris III} D.~L.,  1948, \mn@doi [\apj] {10.1086/145047}, \href
  {https://ui.adsabs.harvard.edu/abs/1948ApJ...108..112H} {108, 112}

\bibitem[\protect\citeauthoryear{Kramers}{Kramers}{1924a}]{Kramers1924a}
Kramers H.~A.,  1924a, Nature, 113, 673

\bibitem[\protect\citeauthoryear{Kramers}{Kramers}{1924b}]{Kramers1924b}
Kramers H.~A.,  1924b, Nature, 114, 310

\bibitem[\protect\citeauthoryear{{Kramers} \& {Heisenberg}}{{Kramers} \&
  {Heisenberg}}{1925}]{Kramers1925}
{Kramers} H.~A.,  {Heisenberg} W.,  1925, \mn@doi [Zeitschrift fur Physik]
  {10.1007/BF02980624}, \href
  {http://adsabs.harvard.edu/abs/1925ZPhy...31..681K} {31, 681}

\bibitem[\protect\citeauthoryear{Landau \& Lifshitz}{Landau \&
  Lifshitz}{1981}]{Landau1981Quantum}
Landau L.~D.,  Lifshitz L.~M.,  1981, Quantum Mechanics Non-Relativistic
  Theory, Third Edition: Volume 3, 3 edn.
Butterworth-Heinemann, \url {http://www.worldcat.org/isbn/0750635398}

\bibitem[\protect\citeauthoryear{Lee}{Lee}{2003}]{lee2003asymmetric}
Lee H.-W.,  2003, The Astrophysical Journal, 594, 637

\bibitem[\protect\citeauthoryear{{Lee}}{{Lee}}{2013}]{Lee2013}
{Lee} H.-W.,  2013, \mn@doi [\apj] {10.1088/0004-637X/772/2/123}, \href
  {http://adsabs.harvard.edu/abs/2013ApJ...772..123L} {772, 123}

\bibitem[\protect\citeauthoryear{{Lee} \& {Lee}}{{Lee} \&
  {Lee}}{1997}]{Lee1997}
{Lee} H.-W.,  {Lee} K.~W.,  1997, \mn@doi [\mnras] {10.1093/mnras/287.1.211},
  \href {https://ui.adsabs.harvard.edu/abs/1997MNRAS.287..211L} {287, 211}

\bibitem[\protect\citeauthoryear{{Mortlock}}{{Mortlock}}{2016}]{Mortlock2016}
{Mortlock} D.,  2016, in {Mesinger} A.,  ed.,  Astrophysics and Space Science
  Library Vol. 423, Understanding the Epoch of Cosmic Reionization: Challenges
  and Progress. p.~187 (\mn@eprint {arXiv} {1511.01107}),
  \mn@doi{10.1007/978-3-319-21957-8_7}

\bibitem[\protect\citeauthoryear{{Morton}}{{Morton}}{2003}]{Morton2003}
{Morton} D.~C.,  2003, \mn@doi [\apjs] {10.1086/377639}, \href
  {https://ui.adsabs.harvard.edu/abs/2003ApJS..149..205M} {149, 205}

\bibitem[\protect\citeauthoryear{{Peebles}}{{Peebles}}{1993}]{Peebles1993}
{Peebles} P.~J.~E.,  1993, {Principles of Physical Cosmology}

\bibitem[\protect\citeauthoryear{Sakurai}{Sakurai}{1967}]{Sakurai1967}
Sakurai J.,  1967, Advanced Quantum Mechanics.
Always learning, Pearson Education, Incorporated, \url
  {https://books.google.co.uk/books?id=lvmSZkzDFt0C}

\bibitem[\protect\citeauthoryear{{Zavarygin}, {Webb}, {Dumont}  \&
  {Riemer-S{\o}rensen}}{{Zavarygin} et~al.}{2018}]{Zavarygin2018}
{Zavarygin} E.~O.,  {Webb} J.~K.,  {Dumont} V.,   {Riemer-S{\o}rensen} S.,
  2018, \mn@doi [\mnras] {10.1093/mnras/sty1003}, \href
  {https://ui.adsabs.harvard.edu/abs/2018MNRAS.477.5536Z} {477, 5536}

\bibitem[\protect\citeauthoryear{van Vleck}{van Vleck}{1924}]{vanVleck1924}
van Vleck J.~H.,  1924, \mn@doi [Phys. Rev.] {10.1103/PhysRev.24.330}, 24, 330

\makeatother
\end{thebibliography}

\section*{Acknowledgements}
CCL thanks the Royal Society for a Newton International Fellowship during the early stages of this work. JKW thanks the John Templeton Foundation for support, the Department of Applied Mathematics and Theoretical Physics and the Institute of Astronomy Cambridge for hospitality and support, and Clare Hall Cambridge for a Visiting Fellowship. In the early stages of this work we had the pleasure of discussing it with Donald Lynden-Bell (who called the work described in this paper ``bread and butter science''!). We are grateful that conversation took place. We also thank Wim Ubachs and Daniel Mortlock for useful comments.

\newpage
\appendix
\section{KHT absorption cross-section near the line centre}
\label{sec:app_case2}

Here is an example to support the validity of equation \eqref{eq:ray6} with the incoming photon frequency, close to the Lyman $\alpha$ line centre, $\nu \rightarrow \nu_{21}$. We first rewrite the summation part of equation \eqref{eq:ray5} as follows,
\begin{eqnarray}
\label{eq:app_ly4}
&& \sum_{n = 2}^\infty A_n^{\mathcal{D}} \left[ \frac{\nu_{\infty}}{ \nu_{n1} \left(1+\frac{by}{c} \right) - \nu + \frac{i \Gamma}{4 \pi}} + \frac{\nu_{\infty}}{ \nu_{n1} \left(1+\frac{by}{c} \right) + \nu} \right] \nonumber \\
&& \qquad = \left\{ \sum_{n=3}^\infty \left[ \frac{\nu_{\infty} A_n^{\mathcal{D}}}{ \nu_{n1} \left(1+\frac{by}{c} \right) - \nu + \frac{i \Gamma}{4 \pi}} + \frac{\nu_{\infty} A_n^{\mathcal{D}}}{ \nu_{n1} \left(1+\frac{by}{c} \right) + \nu} \right]+ \frac{\nu_{\infty} A_2^{\mathcal{D}}}{ \nu_{21} \left(1+\frac{by}{c} \right) + \nu} \right\} + \frac{\nu_{\infty} A_2^{\mathcal{D}}}{ \nu_{21} \left(1+\frac{by}{c} \right) - \nu + \frac{i \Gamma}{4 \pi}}
\end{eqnarray}
The discussion in equation \eqref{eq:taylor_condition} as well as the sentences below shows the condition for Taylor expansion, being $|b \nu_{n1}/c(\nu-\nu_{n1})| \ll 1$ and $|b \nu_{n1}/c(\nu+\nu_{n1})| \ll 1$, and those terms inside the curly brackets, $\{ \cdots \}$, in equation \eqref{eq:app_ly4} obviously meet the requirement, but the last term in equation \eqref{eq:app_ly4} is problematic because $|\nu-\nu_{21}| \rightarrow 0$. To deal with this we follow the same procedure as used for equation \eqref{eq:ray_taylor}.  We perform Taylor series expansions on the terms in the curly brackets in equation \eqref{eq:app_ly4} and on the integration terms in equation \eqref{eq:ray5}. The Rayleigh scattering cross-section then becomes,
\begin{eqnarray}
\label{eq:app_ray_taylor}
\sigma_{Ray}(\nu \rightarrow \nu_{21}) = \sigma_{Th} \int_{-\infty}^\infty dy \frac{e^{-y^2}}{\sqrt{\pi}} \left| \ \frac{\nu_{\infty} A_2^{\mathcal{D}}}{ \nu_{21} \left(1+\frac{by}{c} \right) - \nu + \frac{i \Gamma}{4 \pi}} + \sum_{m=1}^\infty B^\prime_m \left( \frac{by}{c} \right)^{m-1} \ \right|^2 \,,
\end{eqnarray}
where the contribution from the first term inside the absolute bracket in equation \eqref{eq:app_ray_taylor} is absent in $B^\prime_m$, i.e.,
\begin{eqnarray}
\label{eq:app_ray_bm}
B^\prime_m &=& \left\{ \sum_{n =3}^\infty A_n^{\mathcal{D}} \left(\frac{\nu_{\infty}}{\nu_{n1}} \right) \left[ \left( \frac{1}{\nu/\nu_{n1}+1} \right)^m - \left(\frac{1}{\nu/\nu_{n1}-1} \right)^m \right] \right\} \nonumber \\
 && \quad + \int_{0}^{\infty} dk^\prime A_{k^\prime}^{\mathcal{C}} \left(\frac{\nu_{\infty}}{\nu_{k^\prime 1}} \right) \left[ \left( \frac{1}{ \nu/\nu_{k^\prime 1} +1 }  \right)^{m} - \left( \frac{1}{ \nu/\nu_{k^\prime 1} -1} \right)^{m} \right] + A_2^{\mathcal{D}} \left(\frac{\nu_{\infty}}{\nu_{21}} \right) \left( \frac{1}{\nu/\nu_{21}+1} \right)^m \nonumber \\
 &=& \sum_{n=2}^\infty A_n^{\mathcal{D}} \left(\frac{\nu_{\infty}}{\nu_{n1}} \right) \left[ \left( \frac{1}{\nu/\nu_{n1}+1} \right)^m - \left(\frac{1}{\nu/\nu_{n1}-1} \right)^m \right]  \nonumber \\
 && \quad + \int_{0}^{\infty} dk^\prime A_{k^\prime}^{\mathcal{C}} \left(\frac{\nu_{\infty}}{\nu_{k^\prime 1}} \right) \left[ \left( \frac{1}{ \nu/\nu_{k^\prime 1} +1 }  \right)^{m} - \left( \frac{1}{ \nu/\nu_{k^\prime 1} -1} \right)^{m} \right] + A_2^{\mathcal{D}} \left(\frac{\nu_{\infty}}{\nu_{21}} \right) \left(\frac{1}{\nu/\nu_{21}-1} \right)^m \nonumber \\
 &=& B_m + A_2^{\mathcal{D}} \left(\frac{\nu_{\infty}}{\nu_{21}} \right) \left(\frac{1}{\nu/\nu_{21}-1} \right)^m \,.
\end{eqnarray}
The same argument applies to all the line centres, and we have equations \eqref{eq:ray6} and \eqref{eq:ray_b2}, representing the general case for equations \eqref{eq:app_ray_taylor} and \eqref{eq:app_ray_bm}, respectively.

\section{Limiting case -- Recovering a Lorentzian profile from Kramers-Heisenberg}
\label{sec:app_Lorz_recover}

Looking at the form of equations \eqref{eq:ray1} and \eqref{eq:ram1}, the denominator terms involving $(\nu_{n1}-\nu)$ go to zero at the line centre such that only one term in the discrete series remains important,
\begin{eqnarray}
\label{eq:app_ray1}
&& \sigma_{Ray}(\nu) \approx \sigma_{Th} \left| \ |\langle r\rangle_{np,1s}|^2 \left( \frac{\nu \nu_{n1}}{6 \nu_\infty^2 a_0^2} \frac{m_e}{\mu_e} \right) \left( \frac{\nu_\infty}{ \nu_{n1} - \nu  + i \Gamma/4 \pi} \right) \ \right|^2 \,, \\
\label{eq:app_ram1}
&& \sigma_{Ram}^f(\nu) \approx \frac{\sigma_{Th} \nu^\prime}{\nu} \left| \ \langle r\rangle_{f,np} \langle r\rangle_{np,1s} \left( \frac{\nu_{nf} \nu_{n1}}{6 \nu_\infty^2 a_0^2} \frac{m_e}{\mu_e} \right) \left( \frac{\nu_\infty}{\nu_{n1} - \nu + i \Gamma/4 \pi} \right) \ \right|^2 \,.
\end{eqnarray}
Note the superscript and subscript $f$ in \eqref{eq:app_ram1} signifies ``final state'', in the sense described in the main text, not to be confused with $f(n)$, the oscillator strength for each Lyman series transition.

Since only one discrete term dominates when considering the situation $\nu_{n1} \rightarrow \nu$, we are effectively considering a 2-level atom. Hence, we would expect the sum of equations \eqref{eq:app_ray1} and \eqref{eq:app_ram1} to reduce to the semi-classical Lorentzian cross-section, which is
\begin{eqnarray}
\label{eq:lorz}
\sigma_{Lorz}(\nu) = c \sqrt{\frac{3 \pi \sigma_{Th}}{8}} f(n) \frac{\Gamma(n)/4\pi^2}{(\nu-\nu_0)^2 + \left[\Gamma(n)/ 4\pi \right]^2} \,.
\end{eqnarray}

From equation \eqref{eq:app_ray1}, we have
\begin{eqnarray}
\label{eq:app_ray2}
\sigma_{Ray}(\nu) \simeq \sigma_{Th} \frac{\nu_{n1}^2 \nu^2}{36 \nu_\infty^2 a_0^4} \frac{m_e^2}{\mu_e^2} \frac{ | \langle r\rangle_{np,1s}|^4}{\left( \nu_{n1}-\nu \right)^2 + \left( \Gamma /4 \pi \right)^2 } 
\end{eqnarray}
The oscillator strength is given by (see e.g.\ \cite{Corney1977})
\begin{eqnarray}
\label{eq:app_fray}
f(n) = \frac{|\langle r\rangle_{np,1s}|^2}{3 a_0^2} \left( \frac{\nu_{n1}}{\nu_\infty} \right) \left( \frac{m_e}{\mu_e} \right) = \frac{2^8 n^5}{3(n^2-1)^4} \left( \frac{n-1}{n+1} \right)^{2n} \left( \frac{m_e}{\mu_e} \right) \,,
\end{eqnarray}
and the Rayleigh damping constant is
\begin{eqnarray}
\label{eq:app_gam_ray}
\Gamma_{Ray}(n) = \frac{4 \pi \nu^2}{c} \sqrt{\frac{\pi\sigma_{Th}}{6}} \frac{|\langle r\rangle_{np,1s}|^2}{3 a_0^2} \left( \frac{\nu_{n1}}{\nu_\infty} \right) \left( \frac{m_e}{\mu_e} \right) \,.
\end{eqnarray}
Starting with equation \eqref{eq:app_ray2}, we then eliminate $| \langle r\rangle_{np,1s}|^4$ by substitution, using the product of equations \eqref{eq:app_fray} and \eqref{eq:app_gam_ray}. Simplification leads to
\begin{eqnarray}
\label{eq:ray4}
\sigma_{Ray}(\nu) = \sqrt{\frac{3 \pi \sigma_{Th}}{8}} c f(n) \frac{\Gamma_{Ray}(n) /4 \pi^2}{\left(\nu_{n1} - \nu \right) + \left( \Gamma / 4 \pi \right)^2 } \,.
\end{eqnarray}
Equation \eqref{eq:ray4} closely resembles the usual Lorentzian form in equation \eqref{eq:lorz} except we have only so far considered Rayleigh scattering (i.e. angular quantum numbers $\Delta l = \pm 1$). Following the same procedure as above but now for Raman scattering, the damping constants are,
\begin{eqnarray}
\label{eq:gam_ram1}
\Gamma_{Ram}^{s}(n,n_f) = \frac{4 \pi}{c} \left( \frac{\nu^{\prime}}{\nu} \right) \left( \frac{m_e}{\mu_e} \right) \sqrt{\frac{\pi\sigma_{Th}}{6}} \frac{ | \langle r\rangle_{np,n_f s}|^2}{3 a_0^2} \frac{\nu_{n1} \nu_{nf}^2}{\nu_\infty} \,, \\
\label{eq:gam_ram2}
\Gamma_{Ram}^{d}(n,n_f) = \frac{4 \pi}{c} \left( \frac{\nu^{\prime}}{\nu} \right) \left( \frac{m_e}{\mu_e} \right) \sqrt{\frac{\pi\sigma_{Th}}{6}} \frac{ |\langle r\rangle_{np,n_f  d}|^2}{3 a_0^2} \frac{\nu_{n1} \nu_{nf}^2}{\nu_\infty} \,.
\end{eqnarray}
The Raman scattering cross-section is then
\begin{eqnarray}
\label{eq:ram4}
\sigma_{Ram}^{f=n_f,s/d}(\nu) = \sqrt{\frac{3 \pi \sigma_{Th}}{8}} c f(n) \frac{\Gamma_{Ram,s/d}(n, n_f) /4 \pi^2}{\left(\nu_{n1} - \nu \right) + \left( \Gamma / 4 \pi \right)^2 } \,.
\end{eqnarray}
Finally, the Lorentzian profile can be recovered by the sum of equations \eqref{eq:ray4} and \eqref{eq:ram4} near the line centre with
\begin{eqnarray}
\label{eq:app_gamtot}
\Gamma(n) = \Gamma_{Ray}(n) + \sum_{n_f=2}^{n} \left[ \Gamma_{Ram, s}(n,n_f) + \Gamma_{Ram, d}(n,n_f) \right]
\end{eqnarray}
i.e.
\begin{eqnarray}
\label{eq:lorz2}
\sigma_{KH}(\nu \rightarrow \nu_{n1}) &=& \left\{
\sigma_{Ray}(\nu) + \sum_{n_f=2}^{n} \left[ \sigma_{Ram,s}(\nu,n_f,s) + \sigma_{Ram,s}(\nu,n_f,d) \right] \right\}_{\nu \rightarrow \nu_{n1}} \nonumber \\
&\simeq& c \sqrt{\frac{3 \pi \sigma_{Th}}{8}}f \frac{\Gamma(n)/4\pi^2}{(\nu-\nu_0)^2 + (\Gamma(n)/ 4\pi)^2} = \sigma_{Lorz}(\nu)
\end{eqnarray}

\section{Limiting case -- Recovering a Voigt profile from KHT}
\label{sec:app_voigt_recover}

It is clearly important to ensure that we can demonstrate that the KHT profile reduces to the ordinary Voigt profile. We show this by examining the Rayleigh and Raman cross-sections near the line centre (i.e. close to resonance). Equation \eqref{eq:ray6} for Rayleigh scattering contains two terms, the first being the resonance state, the second representing to the interactions between the resonance state and all other discrete levels and continuous terms. The first term therefore corresponds to the Rayleigh contribution to the ordinary Voigt profile (but does not equal the Voigt profile as we need to include the Raman contribution as well). 

The same argument as above but for Raman scattering applied to equation \eqref{eq:ram_voigt1}, that is, the first term represents the resonance (Voigt) term. Therefore we would expect the sum of the first terms (i.e. the resonance contributions without any other discrete or continuous term) in equations \eqref{eq:ray6} and \eqref{eq:ram_voigt1} to be identical to the ordinary Voigt cross-section. i.e. we sum the first terms in equations \eqref{eq:ray_voigt1} and \eqref{eq:ram_voigt1},
\begin{eqnarray}
\sigma_{KHT}(\nu \rightarrow \nu_{n1}) = \sigma_{Ray} + \sum_f \sigma_{Ram}^f \approx \frac{\Gamma_{Ray}(n)}{\Gamma(n)} \sqrt{\frac{3\sigma_{Th}}{8}} \frac{c f_n H(a,u)}{\Delta \nu_{n1}} + \sum_f \frac{\Gamma_{Ram}^{s(d)}(n,n_f)}{\Gamma(n)} \sqrt{\frac{3\sigma_{Th}}{8}} \frac{c f_n H(a,u)}{\Delta \nu_{n1}} \,.
\end{eqnarray}
Using equation \eqref{eq:app_gamtot}, we then have
\begin{eqnarray}
\sigma_{KHT}(\nu \rightarrow \nu_{n1}) \approx \sqrt{\frac{3\sigma_{Th}}{8}} \frac{c f_n H(a,u)}{\Delta \nu_{n1}} \,,
\end{eqnarray}
which recovers the familiar Voigt profile.

\newpage
\section{Oscillator strength and damping constant calculations}
\label{app:appdedix2}

Numerical values for the neutral hydrogen and deuterium damping constants have previously been given by \cite{Morton2003}. However, there are 2 differences between Morton's calculation and the expressions given here.  First, the \cite{Morton2003} values only include the discrete term i.e. they do not include the integral terms in equations \eqref{eq:ray1} and \eqref{eq:ram1}. Second, the \cite{Morton2003} calculation only included Raman scattering for Lyman 1 through 5. Whilst the differences are minor, since some high-precision analyses now make use of higher order lines, we have evaluated equations \eqref{eq:app_fray} and \eqref{eq:app_gamtot} for both hydrogen and deuterium up to Lyman 30 (see table \ref{tab:consts}).

\bigskip

\begin{table}[H]
\centering
\begin{tabular}{c c c c c}
    \toprule
    \midrule
    \multirow{2}[4]{*}{$n$} & \multicolumn{2}{c}{Hydrogen} & \multicolumn{2}{c}{Deuterium}\\ 
    \cmidrule(rl){2-3}
    \cmidrule(rl){4-5}
    & $f(n)$  & $\Gamma(n)$ $(s^{-1})$ & $f(n)$  & $\Gamma(n)$  $(s^{-1})$ \\ 
    \cmidrule(r){1-1}\cmidrule(l){2-3}\cmidrule(l){4-5}
    2 & $4.164 \times 10^{-1} $ & $6.265 \times 10^{8} $ & $4.163 \times 10^{-1} $ & $6.267 \times 10^{8} $ \\
    3 & $7.914 \times 10^{-2} $ & $1.897 \times 10^{8} $ & $7.912 \times 10^{-2} $ & $1.898 \times 10^{8} $ \\
    4 & $2.901 \times 10^{-2} $ & $8.127 \times 10^{7} $ & $2.900 \times 10^{-2} $ & $8.129 \times 10^{7} $ \\
    5 & $1.395 \times 10^{-2} $ & $4.204 \times 10^{7} $ & $1.394 \times 10^{-2} $ & $4.205 \times 10^{7} $ \\
    6 & $7.803 \times 10^{-2} $ & $2.450 \times 10^{7} $ & $7.802 \times 10^{-2} $ & $2.451 \times 10^{7} $ \\
    7 & $4.817 \times 10^{-3} $ & $1.551 \times 10^{7} $ & $4.815 \times 10^{-3} $ & $1.551 \times 10^{7} $ \\
    8 & $3.185 \times 10^{-3} $ & $1.043 \times 10^{7} $ & $3.184 \times 10^{-3} $ & $1.043 \times 10^{7} $ \\
    9 & $2.217 \times 10^{-3} $ & $7.344 \times 10^{6} $ & $2.217 \times 10^{-3} $ & $7.346 \times 10^{6} $ \\
    10 & $1.606 \times 10^{-3} $ & $5.366 \times 10^{6} $ & $1.606 \times 10^{-3} $ & $5.367 \times 10^{6} $ \\
    11 & $1.201 \times 10^{-3} $ & $4.038 \times 10^{6} $ & $1.201 \times 10^{-3} $ & $4.040 \times 10^{6} $ \\
    12 & $9.219 \times 10^{-4} $ & $3.115 \times 10^{6} $ & $9.217 \times 10^{-4} $ & $3.116 \times 10^{6} $ \\
    13 & $7.231 \times 10^{-4} $ & $2.453 \times 10^{6} $ & $7.229 \times 10^{-4} $ & $2.454 \times 10^{6} $ \\
    14 & $5.777 \times 10^{-4} $ & $1.966 \times 10^{6} $ & $5.776 \times 10^{-4} $ & $1.967 \times 10^{6} $ \\
    15 & $4.689 \times 10^{-4} $ & $1.600 \times 10^{6} $ & $4.688 \times 10^{-4} $ & $1.600 \times 10^{6} $ \\
    16 & $3.858 \times 10^{-4} $ & $1.319 \times 10^{6} $ & $3.857 \times 10^{-4} $ & $1.319 \times 10^{6} $ \\
    17 & $3.213 \times 10^{-4} $ & $1.100 \times 10^{6} $ & $3.212 \times 10^{-4} $ & $1.101 \times 10^{6} $ \\
    18 & $2.704 \times 10^{-4} $ & $9.275 \times 10^{5} $ & $2.703 \times 10^{-4} $ & $9.278 \times 10^{5} $ \\
    19 & $2.297 \times 10^{-4} $ & $7.890 \times 10^{5} $ & $2.296 \times 10^{-4} $ & $7.892 \times 10^{5} $ \\
    20 & $1.968 \times 10^{-4} $ & $6.767 \times 10^{5} $ & $1.967 \times 10^{-4} $ & $6.769 \times 10^{5} $ \\
    21 & $1.699 \times 10^{-4} $ & $5.848 \times 10^{5} $ & $1.698 \times 10^{-4} $ & $5.850 \times 10^{5} $ \\
    22 & $1.477 \times 10^{-4} $ & $5.088 \times 10^{5} $ & $1.476 \times 10^{-4} $ & $5.089 \times 10^{5} $ \\
    23 & $1.292 \times 10^{-4} $ & $4.454 \times 10^{5} $ & $1.291 \times 10^{-4} $ & $4.455 \times 10^{5} $ \\
    24 & $1.136 \times 10^{-4} $ & $3.921 \times 10^{5} $ & $1.136 \times 10^{-4} $ & $3.922 \times 10^{5} $ \\
    25 & $1.005 \times 10^{-4} $ & $3.470 \times 10^{5} $ & $1.005 \times 10^{-4} $ & $3.471 \times 10^{5} $ \\
    26 & $8.932 \times 10^{-5} $ & $3.085 \times 10^{5} $ & $8.930 \times 10^{-5} $ & $3.086 \times 10^{5} $ \\
    27 & $7.974 \times 10^{-5} $ & $2.756 \times 10^{5} $ & $7.972 \times 10^{-5} $ & $2.756 \times 10^{5} $ \\
    28 & $7.148 \times 10^{-5} $ & $2.471 \times 10^{5} $ & $7.146 \times 10^{-5} $ & $2.472 \times 10^{5} $ \\
    29 & $6.432 \times 10^{-5} $ & $2.225 \times 10^{5} $ & $6.430 \times 10^{-5} $ & $2.225 \times 10^{5} $ \\
    30 & $5.809 \times 10^{-5} $ & $2.010 \times 10^{5} $ & $5.807 \times 10^{-5} $ & $2.010 \times 10^{5} $ \\
    31 & $5.264 \times 10^{-5} $ & $1.822 \times 10^{5} $ & $5.262 \times 10^{-5} $ & $1.822 \times 10^{5} $ \\
    \midrule
    \bottomrule
\end{tabular}
\caption{.  List of the the oscillator strengths $f$ and the damping constants $\Gamma$ for Hydrogen and Deuterium atoms. Fine structure effects were not taken into account in the calculation.}
\label{tab:consts}
\end{table}

\end{document}